\documentclass[preprint,sort&compress,times,12pt]{elsarticle}
\usepackage{amsmath,graphicx,url}
\newcommand{\bX}{\mathbf{X}}
\newcommand{\bx}{\mathbf{x}}

\begin{document}
\title{Causal Links Between US Economic Sectors}
\author[mas]{Gladys Hui Ting Lee\corref{cor1}}
\author[mas]{Yiting Zhang\corref{cor1}}
\author[mas]{Jian Cheng Wong}
\author[pap]{Manamohan Prusty}
\author[pap]{Siew Ann Cheong\corref{cor2}}
\ead{cheongsa@ntu.edu.sg}

\cortext[cor1]{These authors contributed equally to this work.}
\cortext[cor2]{Corresponding author}

\address[mas]{Division of Mathematical Sciences,
School of Physical and Mathematical Sciences,
Nanyang Technological University,
21 Nanyang Link, Singapore 637371,
Republic of Singapore}

\address[pap]{Division of Physics and Applied Physics,
School of Physical and Mathematical Sciences,
Nanyang Technological University,
21 Nanyang Link, Singapore 637371,
Republic of Singapore}

\begin{abstract}
In this paper, we perform a comparative segmentation and clustering
analysis of the time series for the ten Dow Jones US economic sector
indices between 14 February 2000 and 31 August 2008.  From the
temporal distributions of clustered segments, we find that the US
economy took one and a half years to recover from the
mid-1998-to-mid-2003 financial crisis, but only two months to
completely enter the present financial crisis.  We also find the oil
\& gas and basic materials sectors leading the recovery from the
previous financial crisis, while the consumer goods and utilities
sectors led the descent into the present financial crisis.  On a
macroscopic level, we find sectors going earlier into a crisis emerge
later from it, whereas sectors going later into the crisis emerge
earlier.  On the mesoscopic level, we find leading sectors
experiencing stronger and longer volatility shocks, while trailing
sectors experience weaker and shorter volatility shocks.  In our
shock-by-shock causal-link analysis, we also find shorter delays
between corresponding shocks in more closely related economic sectors.
In addition, our analysis reveals evidences for complex sectorial
structures, as well as nonlinear amplification in the propagating
volatility shocks.  From a perspective relevant to public policy, our
study suggests an endogeneous sectorial dynamics during the mid-2003
economic recovery, in contrast to strong exogeneous driving by Federal
Reserve interest rate cuts during the mid-2007 onset.  Most
interestingly, we find for the sequence of closely spaced interest
rate cuts instituted in 2007/2008, the first few cuts effectively
lowered market volatilities, while the next few cuts
counter-effectively increased market volatilities.  Subsequent cuts
evoked little response from the market.  
\end{abstract}


\maketitle

\section{Introduction}

In a recent paper \cite{Wong2009}, we reported finding the US economy
to be predominantly in a low-volatility phase (which corresponds
roughly to the standard economic growth phase) and a high-volatility
phase (which has a significantly longer duration than the economic
contraction phase it incorporates), when we perform statistical
segmentation and clustering analysis on the Dow Jones Industrial
Average time series between 1997 and 2008.  Both phases are
interrupted by moderate-volatility market correction phases, which
come with two typical durations: 1--2 weeks, and 1.5--2 months.  The
high-volatility phase is also frequently interrupted by
very-high-volatility market crash phases, which can go from 1 day to 3
weeks in duration.  More interestingly, the temporal distribution of
the clustered segments suggests that the US economy made a transition
from the low-volatility phase to the high-volatility phase in mid-1998
(apparently triggered by the July 1997 Asian Financial Crisis), went
back into the low-volatility phase in mid-2003, before entering the
high-volatility phase again in mid-2007 (the current global financial
crisis, apparently triggered by market corrections in the Chinese
markets that started in mid-2006).

Having extracted such a rich and exciting story of the US economy
through segmentation and clustering analysis of a single index time
series, we naturally wondered what we would find if we do a
comparative study of the clustered segments for time series data from
the various US economic sectors.  In particular, we were inclined to
believe that such a comparative analysis offers the potential to
understand causal relationships between different components of the US
economy.  Thus far, many fingers point (with hindsight) to the
complex, poorly managed, poorly regulated interactions between the US
property and financials sectors as the root cause of the present
financial crisis.  However, this stating of what is apparent on the
surface --- less an understanding of the concomitant subtleties
beneath the surface --- may not be enough to help us prevent the next
financial crisis, nor is it likely to show us how to develop effective
mitigation measures should one arise.  If we fully comprehend the
causal linkages between the various economic sectors, we imagine it
would be possible to devise flexible and effective policies that would
target key industries and sectors to accelerate recovery from a
financial crisis, or to soften the impact (or avert altogether) the
onset of a financial crisis.

\begin{figure}[htbp]
\centering
\includegraphics[scale=0.3]{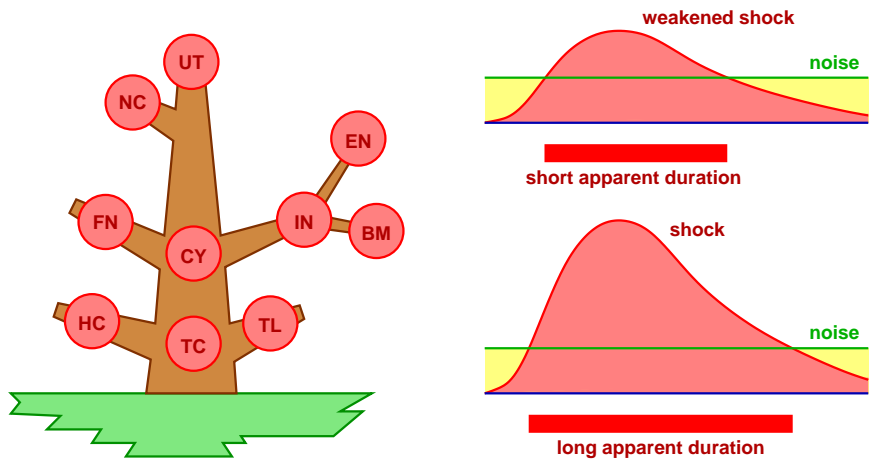}
\caption{(Left) The approximately correct causal tree analogy for
sectorial dynamics within the US economy.  In this analogy, a shock
delivered at the root of the tree will propagate dissipatively upwards
with a finite speed.  TC will experience the strongest shock first,
whereas TL, HC, and CY will experience a weaker shock later, FN and IN
an even weaker shock even later, and finally BM, EN, NC, and UT, on
the outer fringes of the causal tree, experience the weakest shock
latest within the economy.  (Right) Assuming further that the
propagating shock weakens over time, and the signature of the shock
competes with random noises constantly buffeting the sectors, the
identifiable duration of the shock also decreases with distance away
from the root.} 
\label{fig:causaltree} 
\end{figure}

In this paper, we report intriguing results that emerge from our
segmentation and clustering analysis of the ten Dow Jones US economic
sector time series.  The varying temporal distributions of clustered
segments tells us that the last US financial crisis was led to a large
extent by the technologies sector, while the present global financial
crisis was led by the non-cyclical consumer goods (within which we
find the homebuilders and realties) and utilities sectors.
Moreover, we found what we believe might be a generic pattern: the
sector that led the economy into decline recovers last, while the last
sector to succumb to the financial crisis is also the first to recover
when the economy `picks up'.  Other robust statistical signatures
extracted include the leading sectors experiencing stronger and longer
high-volatility shocks, while the trailing sectors experience weaker
and shorter high-volatility shocks.  These observations are consistent
with the causal tree analogy shown in Fig.~\ref{fig:causaltree}.  In
this analogy, exogeneous factors shake the root of the tree, and
branches closest to the root are the first to respond.  As the
influence of the exogeneous shock propagates up the tree, the
amplitudes of the shocks experienced by the sectors decrease in
strength.  Another natural implication of this analogy is that the
delay between successive sectors will be shorter if they are more
closely related.  Because of its apparent utility, we will interpret
our results within the framework of this analogy.

To facilitate easy navigation of our methods and results, this paper
is organized into six sections.  In Section \ref{sect:datamethods}, we
briefly describe the data sets used, as well as the segmentation and
clustering procedures.  In Section \ref{sect:generalfeatures}, we
describe general features observed in the ten temporal distributions
of clustered segments, how we determine the sequence of recovery in
the US economic sectors, as well as the sequence leading into the
current global financial crisis.  In Section \ref{sect:shockbyshock},
we discuss an in-depth shock-by-shock causal-link analysis of the
dynamics of the ten US economic sectors, in the period leading up to
full economic recovery, as well as the start of the Subprime Crisis.
In Section \ref{sect:interestratecuts}, we present results on the
present financial crisis, showing how the economy responded positively
to the first few interest rate cuts, negatively to the next few rate
cuts, and then not at all to subsequent rate cuts.  This suggests that
the Federal Reserve interest rate is not a universal knob that can be
continuously adjusted to fine tune the economy, but can be good
medicine if used sparingly.  Finally, we present our conclusions in
Section \ref{sect:conclusions}.

\section{Data and methods}
\label{sect:datamethods}

\subsection{Data}

Tic-by-tic data for the ten Dow Jones US economic sector indices (see Table
\ref{table:DJUS}) over the period 14 February 2000 to 31 August 2008 were
downloaded from the Taqtic database \cite{Taqtic}.  There are about three
million tic-by-tic records for each index, and the format of this raw data is
shown in Table \ref{table:raw}.  These were processed into half-hourly time
series $\bX_i = (X_{i,1}, \dots, X_{i,t}, \dots, X_{i,N})$, where $i = 1, \dots,
10$ index the various economic sectors according to Table \ref{table:DJUS}, and
$1 \leq t \leq N$ indicate which half-hour within the period of study the
indices are sampled.  For example, in Table \ref{table:raw} we see for BM that
there was a transaction on February 14, 2000 at 14:25:50.259 GMT, at index value
149.92.  The next transaction was on February 14, 2000 at 14:30:29.829 GMT, at
index value 149.93.  Therefore, we take the index value for BM on February 14,
2000 at 14:30:00 GMT, which is the opening time of the New York Stock Exchange
(NYSE), to be 149.92, i.e. the index value of the last transaction before
14:30:00 GMT.  Similarly, the half-hourly index values for 15:00:00 GMT,
15:30:00 GMT, \dots, 20:30:00 GMT, up till the closing time 21:00:00 GMT of the
NYSE, are taken to be the index values of the last transactions before these
half hours.  In the raw data, we also see records of transactions several
minutes after the closing time, and once in a while, we will see transactions an
hour to two hours before the opening time.  Records before the opening time are
corrections made by the NYSE.  These are not real transactions so we ignore such
entries.  Records after the closing time are real transactions, with index
values that can be about 0.1\% different from that of the last transaction
before the official closing time.  This last-minute rush in stock markets is
well known.  We also ignore these records, because their index values will
generally be very close to the index values we assign to the opening hours.
Finally, the list of half hours is also adjusted to take into account daylight
saving.  In the end, we obtained from the tic-by-tic records index values for
31560 half hours for each economic sector.

\begin{table}[htbp]
\centering\footnotesize
\caption{The ten Dow Jones US economic sector indices as defined by
the Industry Classification Benchmark (ICB).  These are float-adjusted
market capitalization weighted sums of variable numbers of component
stocks, introduced on February 14, 2000 to measure the performance of
US stocks in the ten ICB industries.  The makeup of these indices are
reviewed quarterly, and the number of components and float-adjusted
market capitalizations taken from
\protect\url{http://www.djindexes.com/mdsidx/downloads/fact_info/Dow_Jones_US_Indexes_Industry_Indexes_Fact_Sheet.pdf},
and are accurate as of November 30, 2010.  The top components of each
index are shown in Appendix \ref{app:components}.}
\label{table:DJUS}
\vskip .5\baselineskip
\begin{tabular}{cclccc}
\hline
$i$ & symbol & sector & \parbox[c]{2cm}{\raggedright number of 
component stocks} & \parbox[c]{2cm}{\raggedright float-adjusted market
capitalization (billion USD)} & \parbox[c]{2cm}{number of tic-by-tic records} \\
\hline
1 & BM & Basic Materials & 155 & 506.7 & 2,843,033 \\
2 & CY & Consumer Services & 484 & 1,649.1 & 2,937,192 \\
3 & EN & Oil \& Gas & 214 & 1,405.7 & 3,109,893 \\
4 & FN & Financials & 876 & 2,192.5 & 3,086,616 \\
5 & HC & Healthcare & 512 & 1,423.8 & 3,009,245 \\
6 & IN & Industrials & 692 & 1,725.7 & 2,939,937 \\
7 & NC & Consumer Goods & 326 & 1,351.1 & 2,889,067 \\
8 & TC & Technology & 509 & 2,158.1 & 3,222,199 \\
9 & TL & Telecommunications & 44 & 379.5 & 2,908,507 \\
10 & UT & Utilities & 96 & 470.9 & 2,445,898 \\
\hline
\end{tabular}
\end{table}

\begin{table}[htbp]
\centering\footnotesize
\caption{Format of tic-by-tic data downloaded from the Taqtic
database.  In the example shown below for the Dow Jones US economic sector
index for basic materials, the first row is the header of the data
file.  According to this header, the first data column is the Reuters
instrument code (RIC), the second column is the date of the
transactions in MM/DD/YYYY format, the third column is the GMT time of
the transactions in HH:MM:SS.SSS format, the fourth column is the GMT
offset of the exchange, the fifth column refers to the instrument
type, and the sixth column contains the index values.}
\label{table:raw}
\begin{verbatim}
#RIC,Date[G],Time[G],GMT Offset,Type,Price
.DJUSBM,02/14/2000,11:54:20.434,+0,Index,149.92
.DJUSBM,02/14/2000,14:25:50.259,+0,Index,149.92
.DJUSBM,02/14/2000,14:30:29.829,+0,Index,149.93
.DJUSBM,02/14/2000,14:30:57.532,+0,Index,149.92
.DJUSBM,02/14/2000,14:31:28.710,+0,Index,149.93
.DJUSBM,02/14/2000,14:31:57.861,+0,Index,149.94
.DJUSBM,02/14/2000,14:32:15.252,+0,Index,149.93
.DJUSBM,02/14/2000,14:32:36.853,+0,Index,149.94
.DJUSBM,02/14/2000,14:32:59.533,+0,Index,149.95
.DJUSBM,02/14/2000,14:33:13.906,+0,Index,149.98
.DJUSBM,02/14/2000,14:33:30.941,+0,Index,149.97
.DJUSBM,02/14/2000,14:33:43.577,+0,Index,149.98
.DJUSBM,02/14/2000,14:33:58.916,+0,Index,150.02
.DJUSBM,02/14/2000,14:34:13.525,+0,Index,150.05
.DJUSBM,02/14/2000,14:34:30.817,+0,Index,150.15
.DJUSBM,02/14/2000,14:34:43.452,+0,Index,150.16
.DJUSBM,02/14/2000,14:34:58.883,+0,Index,150.33
.DJUSBM,02/14/2000,14:35:13.972,+0,Index,150.34
.DJUSBM,02/14/2000,14:35:31.265,+0,Index,150.39
.DJUSBM,02/14/2000,14:35:45.639,+0,Index,150.29
.DJUSBM,02/14/2000,14:36:02.616,+0,Index,150.38
.DJUSBM,02/14/2000,14:36:20.902,+0,Index,150.41
.DJUSBM,02/14/2000,14:36:40.442,+0,Index,150.32
.DJUSBM,02/14/2000,14:36:49.013,+0,Index,150.22
.DJUSBM,02/14/2000,14:36:59.389,+0,Index,150.19
\end{verbatim}
\end{table}

As explained in Ref.~\cite{Wong2009}, the
half-hourly data frequency allows us to confidently identify
statistically stationary segments as short as a day.  Higher data
frequency was not used, because in a macroeconomic study such as this,
we are not interested in segments shorter than a day.  From the index
time series $\bX_i$, we prepare the log-index movement time series
$\bx_i = (x_{i,1}, \dots, x_{i,t}, \dots, x_{i,N-1})$, where $x_{i, t}
= \log X_{i,t+1} - \log X_{i, t}$, for segmentation based on the
log-normal index movement model described in Ref.~\cite{Wong2009}.
The log-index movement model was chosen because different indices have
different magnitudes, and it is more meaningful to compare their
fractional changes.

\subsection{Segmentation}


Sudden changes in the dynamics of an economy are variously known as
\emph{regime shifts}, \emph{structural breaks}, or \emph{change
points}.  The problem of detecting change points (see for example,
Refs.~\cite{Carlstein1994ChangePointProblems,
Chen2000ParametricStatisticalChangePointAnalysis} is also important in
the fields of image segmentation (see for example,
Refs.~\cite{Pal1993PatternRecognition26p1277,
Zhang1996PatternRecognition29p1335} and biological sequence
segmentation (see for example,
Refs.~\cite{Braun1998StatisticalScience13p142,
Lexa2009StudiesCompIntel204p221}).  In the economics and econometrics
literature, Quandt first considered the problem theoretically in 1958
\cite{Quandt1958JAmStatsAssoc53p873, Quandt1960JAmStatsAssoc55p324},
developing least square estimation procedures used by Huizinga and
Mishkin in their study of regime shifts in US monetary policy
\cite{Huizinga1985NBER1678}.  After putting forth a likelihood test on
regime switching in 1972 \cite{Quandt1972JAmStatsAssoc67p306}, Quandt,
along with Goldfeld, developed in 1973 a Markov switching framework
for estimating regime shifts, and applied it to detect regime shifts
in the US housing market \cite{Goldfeld1973JEconometrics1p3}.  This
Markov switching framework formed the basis of Hamilton's seminal 1989
paper on the US business cycles \cite{Hamilton1989Econometrica57p357,
Driffill1992JEconDynControl16p165}.  There is now a large and growing
economics and econometrics literature on regime shifts and change
points.  Most of these works are based on autoregressive models and
unit-root tests \cite{Andrews1988RevEconStudies55p615,
Andrews1993Econometrica61p821, Bai1994JTimeSerAnal15p453,
Bai1995EconometricTheory11p403, Chong1995EconLett49p351,
Loader1996AnnStats24p1667, Bai1997RevEconStats79p551,
Lumsdaine1997RevEconStats79p212, Bai1998Econometrica66p47,
Lavielle2000JTimeSerAnal21p33, Chong2001EconometricTheory17p87,
Hansen2001JEconPerspec15p117, Zivot2002JBusEconStats20p25,
Bai2003JApplEconometrics18p1, Perron2005JEconometrics129p65,
Guo2006JFinRes29p79, CarrioniSilvestre2009EconometricTheory25p1754},
and only a small number are based on Hamilton's Bayesian approach
\cite{Kim1999RevEconStats81p608}.


In the finance literature, Merton noticed as early as 1976 that stock
returns frequently exhibit discontinuous jumps
\cite{Merton1976JFin31p333, Merton1976JFinEcon3p125}, and extended the
Black-Scholes option pricing framework by incorporating a Poisson jump
process in addition to the standard lognormal returns process.  Ball
and Torous later compared the Black-Scholes-Merton pricing formula to
the actual time series of derivatives, and found only slight
mispricing \cite{Ball1983JFinQuantAnal18p53, Ball1985JFin40p155}.
Jorion, on the other hand, found that jumps in the foreign exchange
market are structurally different from jumps in the stock market
\cite{Jorion1988RevFinStudies1p427}.  In 1986, Poterba and Summers
also noticed persistent changes in the returns variance frequently
accompanied large jumps in the returns
\cite{Poterba1986AmEconRev76p1142}.  Lamoureux and Lastrapes later
associated this persistence of variance with a structural change in
the market \cite{Lamoureux1990JBusEconStats8p225}.  In general,
financial economists are less interested in locating and explaining
change points, and more interested in incorporating the existence of
these temporal features into the prices of derivative instruments.

In contrast to the economics and finance communities, the data mining
communities are very much interested in employing pattern recognition
tools to make predictions.  In pattern-based segmentation schemes,
features within the time series are abstracted into symbols, along the
same spirit as the technical analysis of stock markets
\cite{Murphy1999}.  The time series is then segmented either based on
the relative abundance of symbols, or their context trees
\cite{Chung2002ICDM2002p83, Jiang2007ProcWiCom2007p5609,
Xie2007IntJInfoSysSci3p479, Zhang2007LecNoteCompSci14798p520}.
The focus of the younger econophysics community is again different.
Here, the economy and financial market are seen as complex systems
obeying emergent laws of self organization, and econophysicists
attempt to discover these laws, through studying simple models with
the essential dynamical features, as well as by analyzing
high-frequency market data.  In the time series segmentation works by
Vaglica \emph{et al.} \cite{Vaglica2008PhysRevE77e036110} and T\'oth
\emph{et al.} \cite{Toth2010arXiv10012549}, the aim is to discover
scaling laws governing financial markets, whereas in our previous work
\cite{Wong2009}, we are concerned with what macroeconomic phases are
manifest in a financial market, and what time scales are associated
with the transition from one macroeconomic phase to another.

In this paper, we would like to learn more about the dynamics of the US economy,
by segmenting the time series of the ten Dow Jones US economic sector indices.
To do this, we assume that each economic sector time series $\bx_i$ consist of
$M_i$ segments, and that within segment $m_i$, the log-index movements
$x^{m_i}_{i, t}$ are normally distributed, with constant mean $\mu_{i, m_i}$ and
constant variance $\sigma_{i, m_i}^2$.  The unknown segment boundaries $t_{i,
m_i}$, which separates segments $m_i$ and $m_i + 1$, are determined through time
series segmentation, using the recursive entropic scheme introduced by
Bernaola-Galv\'an \emph{et al}.  \cite{BernaolaGalvan1996PhysicalReviewE53p5181,
RomanRoldan1998PhysicalReviewLetters80p1344}.  In this segmentation scheme, we
start with the time series $\bx = (x_{1}, \dots, x_{t},$ $x_{t+1}, \dots,
x_{n})$, and compute the Jensen-Shannon divergence
\cite{Lin1991IEEETransactionsonInformationTheory37p145}
\begin{equation}
\Delta(t) = \ln\frac{L_2(t)}{L_1},
\end{equation}
where within the log-normal index movement model,
\begin{equation}
L_1 = \prod_{s=1}^n \frac{1}{\sqrt{2\pi\sigma^2}}
\exp\left[-\frac{(x_{s} - \mu)^2}{2\sigma^2}\right]
\end{equation}
is the likelihood that $\bx$ is generated probabilistically by a
single Gaussian model with mean $\mu$ and variance $\sigma^2$, and 
\begin{equation}
L_2(t) = \prod_{s=1}^t \frac{1}{\sqrt{2\pi\sigma_L^2}}
\exp\left[-\frac{(x_s - \mu_L)^2}{2\sigma_L^2}\right]
\prod_{s=t+1}^n \frac{1}{\sqrt{2\pi\sigma_R^2}}
\exp\left[-\frac{(x_s - \mu_R)^2}{2\sigma_R^2}\right]
\end{equation}
is the likelihood that $\bx$ is generated by two statistically
distinct models: the left segment $\bx_L = (x_1, \dots, x_t)$ by a
Gaussian model with mean $\mu_L$ and variance $\sigma_L^2$, and the
right segment $\bx_R = (x_{t+1}, \dots, x_n)$ by a Gaussian model with
mean $\mu_R$ and variance $\sigma_R^2$.  In terms of the maximum
likelihood estimates $\hat{\mu}, \hat{\mu}_L, \hat{\mu}_R$ and
$\hat{\sigma}^2, \hat{\sigma}_L^2, \hat{\sigma}_R^2$, the
Jensen-Shannon divergence $\Delta(t)$, which measures how much better
a two-segment model fits the time series data compared to a one-segment
model, simplifies to
\begin{equation}\label{eqn:simpleJS}
\Delta(t) = n\ln\hat{\sigma} - t \ln \hat{\sigma}_L - (n - t) \ln
\hat{\sigma}_R + \frac{1}{2} \geq 0.
\end{equation}

\begin{figure}[htbp]
\centering
\includegraphics[scale=0.5,clip=true]{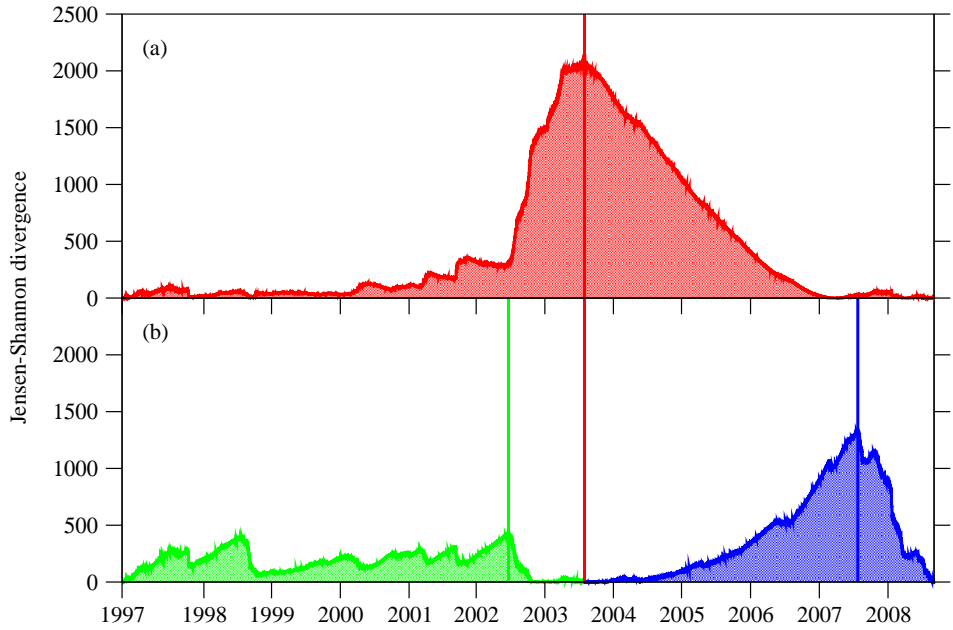}
\caption{The (a) first and (b) second stages of the recursive
segmentation of the half-hourly Dow Jones Industrial Average time
series between January 1, 1997 and August 31, 2008.  In the first
stage of the recursive segmentation, the Jensen-Shannon divergence
$\Delta(t)$ is computed over the entire time series, and mid-2003 is
found to be the optimum time point for the first segment boundary.  In
the second stage of the recursive segmentation, $\Delta(t)$ is
computed independently over the left and right segments.  The optimum
time points for the second and third segment boundaries are found to
be mid-2002 and mid-2007 respectively.}
\label{fig:JSseg}
\end{figure}

We then scan through all possible times $t$, as shown in
Fig.~\ref{fig:JSseg}(a), and place a cut at $t^*$, for which the Jensen-Shannon
divergence
\begin{equation}
\Delta^* = \Delta(t^*) = \max_t \Delta(t)
\end{equation}
is maximized, to break the time series $\bx = (x_1, \dots, x_n)$ into two
statistically most distinct segments $\bx_L^* = (x_1, \dots, x_{t^*})$ and
$\bx_R^* = (x_{t^* + 1}, \dots, x_n)$.  In Fig.~\ref{fig:JSseg}, the half-hourly
time series shown is that of the Dow Jones Industrial Average between January 1,
1997 and August 31, 2008.  The first segment boundary identified by this
one-to-two segmentation procedure is at $t_1^* \approx$ mid-2003, since this is
where $\Delta(t)$ peaks, when it is computed over the entire time series.  The
Jensen-Shannon divergence value associated with this peak is $\Delta_1^* \approx
2000$.  Recalling that $\Delta(t) = \ln L_2(t)/L_1$, this means that at this
point in time, the two-segment likelihood is $e^{2000}$ larger than the
one-segment likelihood.  Therefore, when benchmarked against the model of one
stationary segment for the entire time series, we know that the left segment
from January 1997 to mid-2003 is statistically very dissimilar to the right
segment from mid-2003 to August 2008.  Given such a large disparity between the
one-segment and two-segment likelihoods, it is clear that the mid-2003 segment
boundary is highly significant statistically.  When we use Eq.~\eqref{eqn:dDmax}
calculate the maximum error that could arise in the Jensen-Shannon divergence of
the whole time series, which has 31560 points, we find $\delta\Delta_{\max} =
52$.  This also suggests that the mid-2003 segment boundary, with $\Delta_1^*
\approx 2000$, is statistically highly significant.  In fact, this time point
corresponds to the start of the four-year growth phase of the US economy from
mid-2003 to mid-2007.

\begin{figure}[htb]
\centering
\includegraphics[scale=0.5,clip=true]{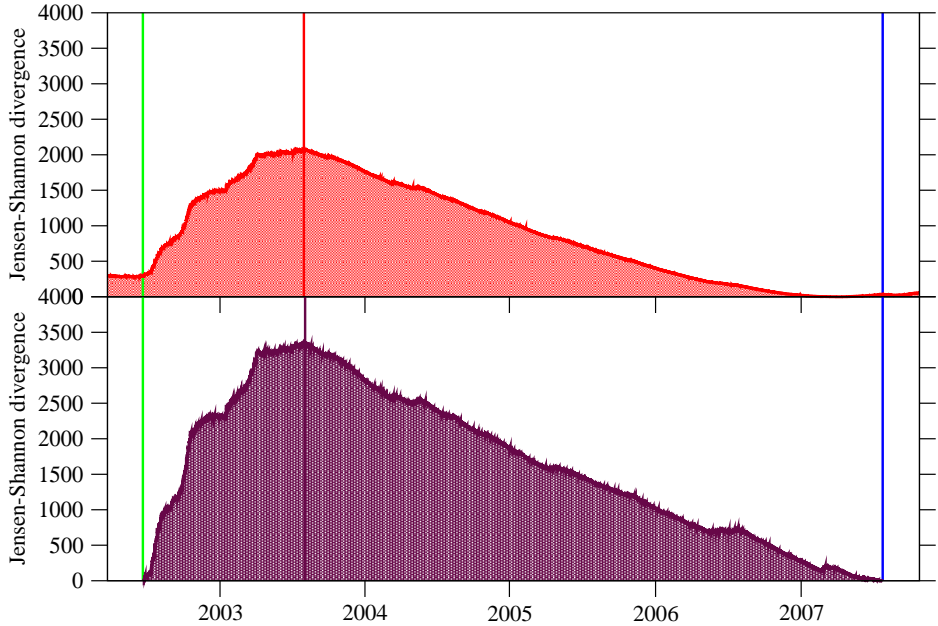}
\caption{The Jensen-Shannon divergence $\Delta(t)$ of the Dow Jones
Industrial Average time series computed in (a) the first stage of the
recursive segmentation over the entire half-hourly time series from
January 1997 to August 2008.  The optimum segment boundary was
identified to be at $t_1^* \approx$ mid-2003.  After the second stage
of the recursive segmentation, two new segment boundaries were
discovered at $t_2^* \approx$ mid-2002 and $t_3^* \approx$ mid-2007.
When $\Delta(t)$ was recomputed (b) over the interval $(t_2^*,
t_3^*)$, the optimum position for $t_1^*$ is now shifted slightly to
${t'_1}^*$.}
\label{fig:segopt}
\end{figure}

To apply this one-into-two segmentation scheme recursively to obtain shorter and
shorter segments, we compute $\Delta(t)$ separately for $\bx^*_L$ and $\bx^*_R$,
as shown in Fig.~\ref{fig:JSseg}(b).  For $\bx_L^*$, we find the optimum segment
boundary to be at $t^*_2 \approx$ mid-2002, whereas for $\bx_R^*$, the optimum
segment boundary is at $t_3^* \approx$ mid-2007.  After this second stage
segmentation, we now have four segments, separated by two new and one old
segment boundary.  The two new segment boundaries have $\Delta_2^* \approx 500$
(mid-2002, with $\delta\Delta_{\max} \approx 40$) and $\Delta_3^* \approx 1500$
(mid-2007, with $\delta\Delta_{\max} \approx 35$) respectively.  Both are thus
less significant than the first segment boundary discovered by the segmentation
procedure.  However, they remain highly significant statistically.  In fact,
$t_2^*$ corresponds to the mid-2002 Dow Jones low, and $t_3^*$ corresponds to
the July 2007 start of the Subprime Crisis.  Before we move on to the third
stage of the recursive segmentation, we need to ensure that the position of
$t_1^*$ remains optimum.  This is necessary, because $t_1^*$ was initially
identified using $\Delta(t)$ computed the entire time series.  But now that we
know the positions of $t_2^*$ and $t_3^*$, we should compute $\Delta(t)$ over
the interval $(t_2^*, t_3^*)$ only to locate $t_1^*$.  When we do this, the
optimum position for $t_1^*$ must be shifted slightly to ${t'_1}^*$, as shown in
Fig.~\ref{fig:segopt}.  Since $t_1^*$ has been moved, we must also check the
optimalities of $t_2^*$ and $t_3^*$, by computing $\Delta(t)$ over the intervals
$(1, {t'_1}^*)$ and $({t'_1}^*, n)$.  At each stage of the recursive
segmentation, this first-order optimization \cite{CheongIRJSS} must be done
iteratively for all segment boundaries, until they have all converged onto their
optimum positions.

\begin{figure}[htbp]
\centering
\includegraphics[scale=0.5,clip=true]{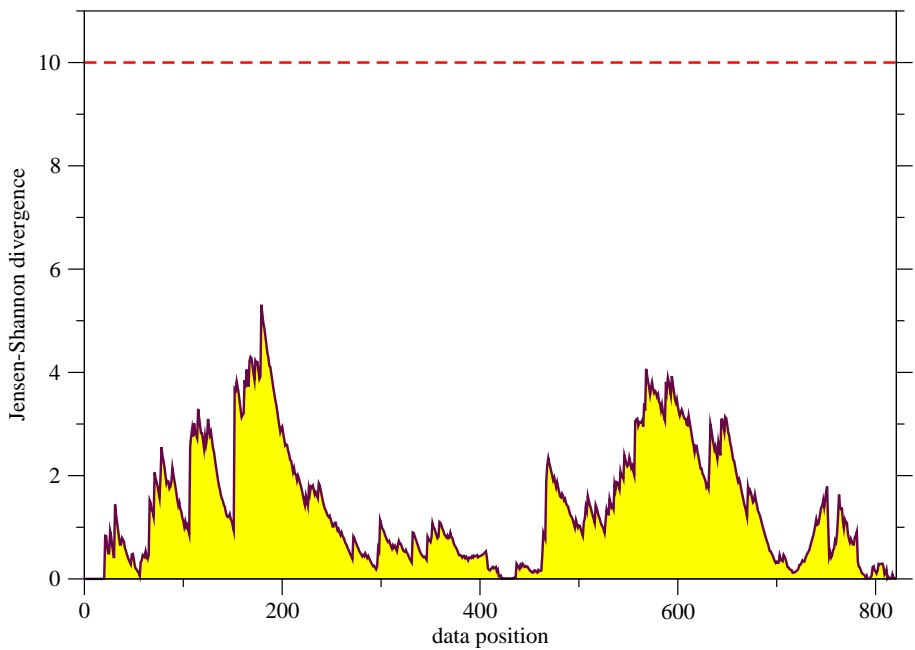}
\caption{The Jensen-Shannon divergence $\Delta(t)$ of a moderate-length (821
half hours, or equivalently, about 55 trading days) times series segment of the
Dow Jones Industrial Average.  Contrast this to Fig.~\ref{fig:JSgaussian}, where
$\Delta(t)$ is obtained from a 821-point time series generated by a stationary
Gaussian process.  In this example, the peak $\Delta_{\max} \approx 5.5$ at the
data position of about 200 is statistically too weak to justify the introduction
of a new boundary within this segment, according to both our empirical cutoff of
$\Delta_0 = 10$, and also the error of $\delta\Delta = 7.4$ calculated from
Eq.~\eqref{eqn:dD}.}
\label{fig:terminate}
\end{figure}

\newpage
From Fig.~\ref{fig:segopt}, we also see that ${\Delta'_1}^* \approx 3500$
obtained over the interval $(t_2^*, t_3^*)$ is actually larger than $\Delta_1^*
\approx 2000$ obtained over the entire time series.  This is in spite of the
maximum error in computing the Jensen-Shannon divergence falling from
$\delta\Delta_{\max} \approx 50$ over the whole time series, to
$\delta\Delta_{\max} \approx 35$ over $(t_2^*, t_3^*)$.  Therefore, $\Delta_2^*
\approx 500$ and $\Delta_3^* \approx 1500$ are smaller than $\Delta_1^*$ not
because they are computed over shorter segments, but because the segment
boundaries $t_2^*$ and $t_3^*$ are in fact statistically less significant than
the segment boundary $t_1^*$.  Nevertheless, as the optimized recursive
segmentation progresses, all the highly significant segment boundaries would
have been discovered, and we start discovering less and less significant segment
boundaries.  When this happens, the Jensen-Shannon divergence of the newly
discovered segment boundaries will become smaller and smaller.  When $\Delta(t)$
starts looking like that shown in Fig.~\ref{fig:terminate}, we can no longer
identify any statistically significant boundary within the segment.  Such
segments should therefore not be segmented any further.

\newpage

At this point, we find it necessary to address the important question of how we
decide whether a time series is statistically stationary or nonstationary.  In
the statistics and econometrics literatures, a time series that fails a
unit-root test when it is fitted to an autoregressive model can plausibly be
regarded as statistically nonstationary.  However, in its most general terms
this question is not well posed: no matter how nonstationary a given time series
looks, it can always be fitted to a stationary stochastic process.  Similarly,
it is also possible for a stationary model to produce a seemingly nonstationary
time series, or for a nonstationary model to produce a seemingly stationary
time series.  A stationary model is the simplest model for any given time
series.  However, if the likelihood for observing the given time series is very
low, then it is not better than a more complex nonstationary model which
reproduces the given time series with much higher likelihood.  Therefore, the
more meaningful question to ask is if a given time series can be more profitably
modeled by a stationary model or by a nonstationary model.  Based on our
discussions above, the answer to this model selection problem is very clear in
the initial stages of the recursive segmentation, when the two-segment
likelihoods are so much larger than the one-segment likelihoods.  More
importantly, the peak Jensen-Shannon divergence is also very much larger than
the amplitude of the point-to-point fluctuations in $\Delta(t)$ across the time
series. 

\begin{figure}[htb]
\centering
\includegraphics[scale=0.5,clip=true]{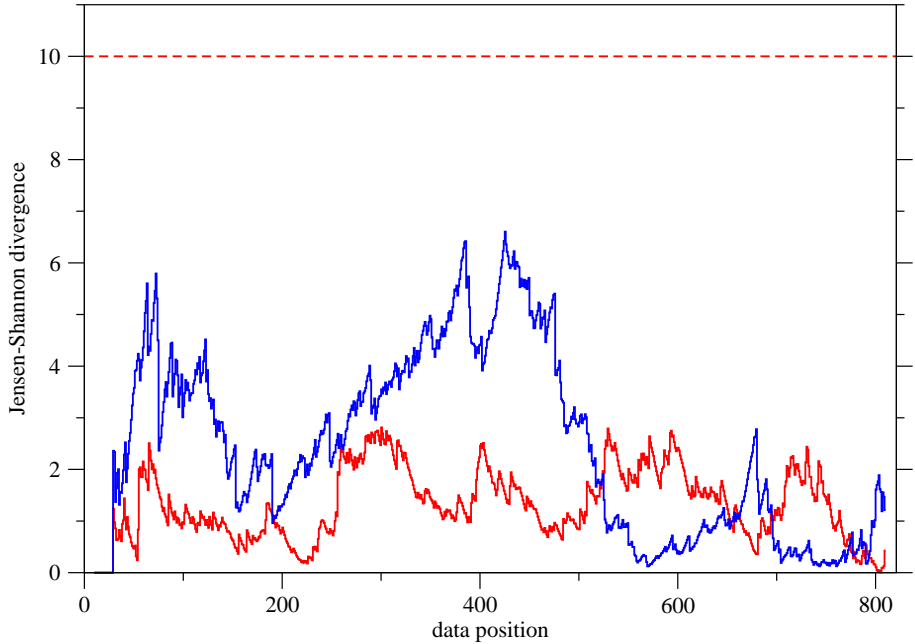}
\caption{The Jensen-Shannon divergence $\Delta(t)$ of two 821-point time series
generated by a stationary Gaussian process.  For concreteness, we chose $\mu =
0$ and $\sigma = 1$ for the stationary Gaussian process, even though the
Jensen-Shannon divergences depends only on the how inhomogeneous the time series
is.  Fairly large $\Delta^*$ can be observed, but these in general do not exceed
$\Delta_0 = 10$.  The character of the point-to-point fluctuations in
$\Delta(t)$ is also different from that observed in Fig.~\ref{fig:terminate}.}
\label{fig:JSgaussian}
\end{figure}

At the late stages of recursive segmentation, most plots of $\Delta(t)$ look
like that shown in Fig.~\ref{fig:terminate}.  The peak $\Delta^*$ is now small,
and not that much larger than the point-to-point fluctuations in $\Delta(t)$.
Discounting structures seen in the larger-scale fluctuations,
$\Delta(t)$ now resembles those of artificial time series (shown in
Fig.~\ref{fig:JSgaussian}) generated by stationary Gaussian processes.  The
segment shown in this example should therefore not be further segmented.  As far
as we are aware of, there are three statistical frameworks for terminating the
recursive segmentation in the literature.  In the original approach by
Bernaola-Galv\'an and coworkers \cite{BernaolaGalvan1996PhysicalReviewE53p5181,
RomanRoldan1998PhysicalReviewLetters80p1344}, the divergence maximum of a new
segment boundary is tested for statistical significance against a $\chi^2$
distribution whose degree of freedom depends on the length of the segment to be
subdivided.  Recursive segmentation terminates when no new segment boundaries
more significant than the chosen confidence level can be found.  In the second
approach \cite{Li2001PhysicalReviewLetters86p5815, Li2001ProcRECOMB01p204},  a
segment is subdivided if the information criterion of its best two-segment model
exceeds that of its one-segment model.  Recursive segmentation terminates when
further segmentation does not explain the data better.  In the third approach
\cite{CheongIRJSS}, we compare the Jensen-Shannon divergence $\Delta(t)$ against
a coarse-grained divergence $\tilde{\Delta}(t)$ of the segment to be subdivided,
to compute the total strength of point-to-point fluctuations in $\Delta(t)$.
Recursive segmentation terminates when the area under $\tilde{\Delta}(t)$ falls
below the desired signal-to-noise ratio.

All the most statistically significant segment boundaries will be discovered by
recursive segmentation using any of the three termination criteria.  Based on
the experience in our previous work \cite{Wong2009}, these most statistically
significant segment boundaries are also discovered if we terminate the recursive
segmentation when no new optimized segment boundaries with Jensen-Shannon
divergence greater than a cutoff of $\Delta_0 = 10$ are found.  This choice of
cutoff is consistent with the standard errors $\delta\Delta$ calculated using
Eq.~\eqref{eqn:dD} and the eventual sizes of the segments, although it sometimes
result in long segments whose internal segment structures are masked by their
context \cite{CheongCSP}.  For these long segments, we progressively lower the
cutoff $\Delta_0$ until a segment boundary with strength $\Delta > 10$ appears.
The final segmentation then consists of segment boundaries discovered through
the automated recursive segmentation, as well as segment boundaries discovered
through progressive refinement of overly long segments.

Before we move on to describe how segments are grouped into a small number of
classes based on their statistical similarities, let us also discuss how the
segmentation procedure will perform if the volatility $\sigma(t) = \sigma_1(t) +
\sigma_2(t)$ consists not only of a deterministic part $\sigma_1(t)$, but also
receives contribution from a stochastic part $\sigma_2(t)$.  For our
segmentation procedure to work, $\sigma_1(t)$ and $\sigma_2(t) =
\beta(t)\,dZ(t)$ must change abruptly from one segment to the next.  Here,
$\beta(t)$ is a deterministic parameter, and $dZ(t)$ is a stochastic variable
drawn from any standard distribution.  Whether it is $\sigma_1(t)$, $\beta(t)$,
or both, that is undergoing a sudden transition, this change point will be
detected if it is statistically significant (and thus likely to be economically
meaningful).  In this sense, we do not need to specifically worry about any
stochastic contributions to the segment volatilities.

\subsection{Clustering}

After the time series segmentation is completed, 
we end up with between 100 and 150 segments for each economic sector
index.  For each time series, a segment is statistically distinct from
the segment before it as well as the segment after it.  However,
distant segments can be statistically similar to each other.  In this
way, we expect the large number of segments may actually represent a
smaller number of segment \emph{types} or \emph{classes}.  This was
the case when Azad \emph{et al.} segmented the human chromosome 22,
and found that the 248 segments can be classified into 53 segment
types \cite{Azad2002PhysRevE66e031913}.  In fact, there is good reason
to believe that the time series segments we obtained can actually be
organized into a small number of classes, each representing a
macroeconomic phase.

The procedure of assigning a large number of objects into a smaller
number of collections, such that within each collection, the objects
are more similar to each other than they are with objects from another
collection, is known as \emph{clustering} or \emph{classification}
(see for example, the books by Mirkin
\cite{Mirkin1996MathematicalClassificationClustering} and by Halgamuge
and Wang
\cite{Halgamuge2005ClassificationClusteringKnowledgeDiscovery}, or the
review by Jain \cite{Jain1999}).  Clustering algorithms can be broadly
classified as \emph{partitional} or \emph{hierarchical}.  In the
$k$-means algorithm
\cite{MacQueen1967Proc5thBerkeleySympMathStatsProb1p281,
Lloyd1982IEEETransInforTheor28p129}, which is the representative
algorithm for partitional clustering, we decide beforehand that there
are $k$ clusters, and assign each data point to a cluster, such that
the sum of square deviations to the $k$ means is minimized by varying
the centers of the clusters as well as the cluster assignment.  In
single linkage clustering \cite{Sneath1957JGenMicrobiol17p201,
Johnson1967Psychometrika32p241}, by far the most popular hierarchical
clustering algorithm, small clusters are progressively merged into
larger clusters, by first merging clusters that are closest together.
In this clustering algorithm, the `distance' between two clusters is
given by the smallest `distance' between their constituents.

Clustering of different periods within a financial time series has
been previously investigated by van Wijk \emph{et al.}
\cite{vanWijk1999ProcInfoVisualp4} and Fu \emph{et al}
\cite{Fu2004ProcIntConfDataMiningp5}, with the goal of discovering
patterns that can be used for doing prediction.  In this paper, we
perform hierarchical agglomerative clustering of the time series
segments to organize them into different macroeconomic phases.  We do
this for each US economic sector index independently, because the same
macroeconomic phase may exhibit different statistical characteristics
in different indices.  Also, as we are interested in discovering
macroeconomic phases, we use the complete link algorithm
\cite{Baker1972RevEduRes42p345}, favored by social scientists for
producing compact clusters with the maximum internal homogenuity.  We
do not use the single link algorithm, which is more meaningful in the
biological sciences because it corresponds more closely with the
nature of evolutionary changes, since it tends to produce loose and
elongated clusters \cite{Baker1974JAmStatsAssoc69p440}.

\begin{figure}[htbp]
\centering
\includegraphics[scale=0.35]{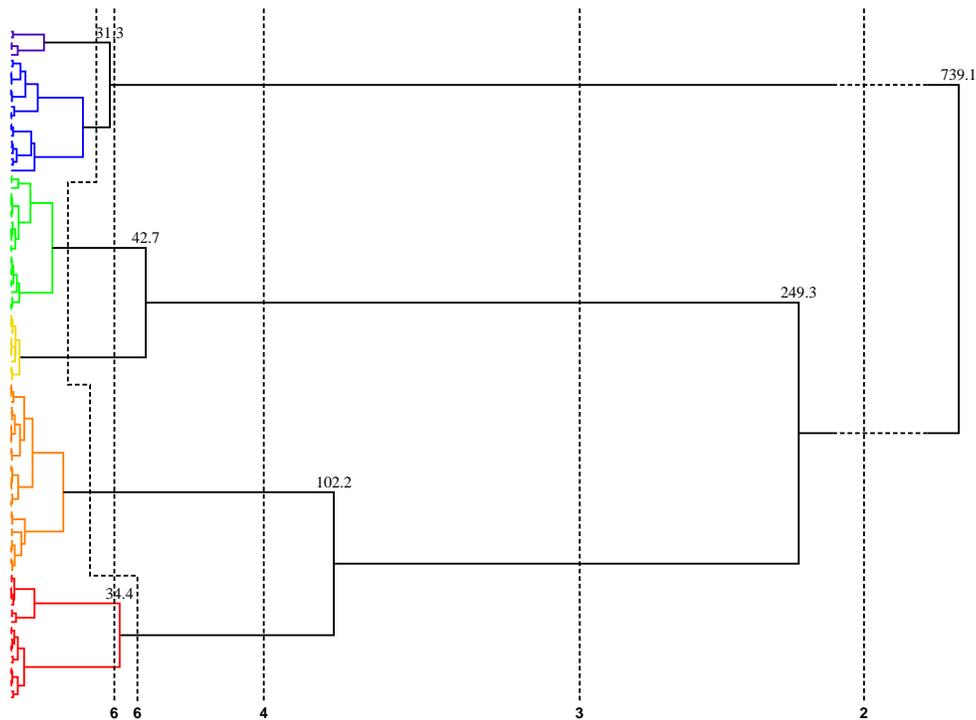}
\caption{The complete-link hierarchical clustering tree for the time
series segments of the Dow Jones Industrial Average between January
1997 and August 2008.  In this tree, we show the Jensen-Shannon
divergence values at which the top branches diverge.  We also show how
uniform thresholds can be selected to break the tree into two, three,
four, or six clusters.  Finally, we show how individual thresholds can
be selected to obtain the six clusters reported in
Ref.~\cite{Wong2009}, which are colored in increasing order of market
volatility as deep blue, blue, green, yellow, orange, and red.}
\label{fig:cltree}
\end{figure}

There are many ways to extract clusters from a hierarchical clustering
tree.  For example, in the hierachical clustering tree of time series
segments of the Dow Jones Industrial Average shown in
Fig.~\ref{fig:cltree}, we can group the segments into two clusters, if
we choose the threshold statistical distance, measured by the
Jensen-Shannon divergence between segments, to be $249.3 < \Delta <
739.1$, or three clusters, if we choose the threshold statistical
distance to be $102.2 < \Delta < 249.3$.  We can also group the
segments into six clusters (different from the ones identified by
colors), if we choose the threshold statistical distance to be $31.3 <
\Delta < 34.4$.  However, unlike time series segmentation, the
statistical criterion for a meaningful clustering is not significance,
but robustness.  A robust cluster is one whose composition does not
change over a broad range of threshold statistical distances.  At a
higher level, a robust clustering is one in which the number of
clusters does not change over a broad range of threshold statistical
distances.  Once we understand this different statistical concern, we
can even work with different thresholds for different clusters.  In
Fig.~\ref{fig:cltree}, we make use of this flexibility to identify
the six colored clusters for the Dow Jones Industrial Average.  

In the same way, we analyzed the hierarchical complete-link clustering
trees obtained for the ten US economic sector indices, and selected
between four to six coarse-grained clusters for each index.  In
general, we choose to work with between four and six clusters, instead
of fewer or more, because we want to map the clusters to the four
macroeconomic phases, growth, crisis, correction, and crash,
identified by economists.  This mapping between clusters and
macroeconomic phases, as well as the color scheme used in all the time
series plots in this paper, is shown in Table \ref{table:heatmap}.  We
have shown in Ref.~\cite{Wong2009} that this association between
clusters and macroeconomic phases is reasonable, by correctly
identifying the start and end of crises and growths of the US economy
from 1997 to 2008.  In this paper, we hope that similar analysis based
on the temporal distributions of clustered segments for the ten US
economic sectors, presented in Sections \ref{sect:generalfeatures},
\ref{sect:shockbyshock}, and \ref{sect:interestratecuts}, will shed
more light on the sectorial dynamics within the US economy.

\begin{table}[htbp]
\centering\footnotesize
\caption{Heat-map-like color scheme for the different volatility clusters, and
the macroeconomic phases they correspond to.  The crisis phase, which consists
of the high-volatility (yellow) and very-high-volatility (orange) clusters, is
significantly longer than the economic contraction phase accepted by economists.
In fact, economic contraction, as determined by successive quarters of
contraction in the GDP, typically occurs at the end of a crisis phase.  Also
shown are the average standard deviation in each phase for the various economic
sectors.  Ideally, if we believe there are only four distinct macroeconomic
phases, we can use various methods found in the broader statistics literature
(for instance, by building and calibrating a hidden Markov model
\cite{Churchill1989BullMathBiol51p79, Churchill1992CompChem16p107,
Bize1999ProcRECOMB, Peshkin1999Bioinformatics15p980, Boys2000ApplStats49p269,
Nicolas2002NucleicAcidsRes30p1418, Boys2004Biometrics60p573}) to determine the
volatility distributions in each of these four macroeconomic phases.  Once these
volatility distributions are discovered, a given segment volatility can then be
assigned to a macroeconomic phase using likelihood-based measures.  However, we
would like to discover for ourselves the numbers of robust volatility classes,
without assuming that they are the same for all economic sectors.  In other
words, we let the high-frequency time series data tell us what volatility
classification is most `natural' for each of the ten DJUS economic sector
indices.  We find that the segment volatilities for BM, HC, IN, NC, TC, TL, UT
can be most naturally organized into five clusters, whereas those for CY, EN, FN
can be most naturally organized into six clusters (the sixth being an
extremely-low-volatility cluster).  During the segment clustering, we also did
not assume the same ranges of volatilities across all indices for the same
cluster.  But as we can see, within each macroeconomic phase, the average
volatilities discovered by the clustering procedure are fairly consistent
throughout most sectors.  The exceptions are HC and TL, which have consistently
lower volatilities.  We could have introduce a seventh cluster with volatility
$\sigma \approx 0.008$, and reclassified all the clusters.  This will produce a
deterministic mapping between volatility and color, but we choose not to, so as
to achieve maximum visual contrast with the present color scheme.}

\label{table:heatmap}
\vskip .5\baselineskip
\begin{tabular}{|c|c|c|c|c|c|c|}
\hline
\emph{volatility} & 
extremely low & low & moderate & high & very high & extremely high \\
\hline
\emph{color} & black & blue & green & yellow & orange & red \\
\hline
\emph{phase} & \multicolumn{2}{|c|}{growth} & correction &
\multicolumn{2}{|c|}{crisis} & crash \\
\hline
BM & - & 0.0016 & 0.0037 & 0.0046 & 0.0069 & 0.0146 \\
CY & 0.0005 & 0.0015 & 0.0023 & 0.0031 & 0.0053 & 0.0121 \\
EN & 0.0010 & 0.0014 & 0.0027 & 0.0037 & 0.0058 & 0.0152 \\
FN & 0.0007 & 0.0016 & 0.0024 & 0.0039 & 0.0058 & 0.0134 \\
HC & - & 0.0006 & 0.0016 & 0.0023 & 0.0041 & 0.0076 \\
IN & - & 0.0013 & 0.0022 & 0.0035 & 0.0056 & 0.0140 \\
NC & - & 0.0009 & 0.0015 & 0.0022 & 0.0034 & 0.0085 \\
TC & - & 0.0019 & 0.0030 & 0.0042 & 0.0082 & 0.0121 \\
TL & - & 0.0008 & 0.0018 & 0.0024 & 0.0033 & 0.0078 \\
UT & - & 0.0014 & 0.0023 & 0.0030 & 0.0038 & 0.0088 \\
\hline
\end{tabular}
\end{table}

\section{Temporal distributions of clustered segments: general
features}
\label{sect:generalfeatures}

As expected, the time series of different economic sectors exhibit different
temporal distributions of clustered segments (see Appendix
\ref{app:listsegments} for complete listings of time series segments found for
the ten DJUS indices).  However, based on the distributions of high-volatility
segments (which are dominant during financial crises) and low-volatility
segments (which are dominant during the economic expansion phase), we see that
all economic sectors went into the high-volatility phase during the previous
financial crisis, reverted to the low-volatility phase, and then entered the
high-volatility phase again during the present global financial crisis.  Our
main interest lies in whether we can draw meaningful inferences on the causal
relationships between the various US economic sectors, by studying these
consistent time series features that emerge during recovery from the previous
financial crisis, and the onset of the present financial crisis.

\subsection{Recovery from mid-1998 to mid-2003 financial crisis}

For any given US economic sector, its time series segment boundaries are not
equally significant.  Some segment boundaries have large $\Delta^*$, and are
thus highly significant statistically.  Other segment boundaries have $\Delta^*$
just above our cutoff of $\Delta_0 = 10$, and are thus less significant
statistically.  When we cluster these time series segments, we not only group
temporally distant segments which are statistically similar, we also group
adjacent segments separated by statistically weaker boundaries.  In our temporal
distribution of clustered segments plot, adjacent segments assigned to the same
cluster will be mapped to the same color.  Conversely, adjacent segments which
are colored differently must have been assigned to different clusters, because
they have highly dissimilar statistics, and hence the boundary separating them
is highly significant statistically.  For example, as shown in Appendix
\ref{app:listsegments}, the extremely-high-volatility segment $m = 9$ of BM
($\sigma = 0.006626 \pm 0.001210$) is flanked by the low-volatility segments $m
= 8$ ($\sigma = 0.001186 \pm 0.000040$) and $m = 10$ ($\sigma = 0.001354 \pm
0.000062$).  The Jensen-Shannon divergence of the boundary between $m = 8$ and
$m = 9$ is $\Delta = 103.0 \pm 2.7$, whereas the Jensen-Shannon divergence of
the boundary between $m = 9$ and $m = 10$ is $\Delta = 35.0 \pm 2.5$.
Furthermore, because the clusters we identified from the hierachical clustering
tree are highly robust, the set of time points where the color in the temporal
distribution change is also highly robust.  We therefore design our feature
extraction methodology around this set of statistically significant and robust
segment boundaries.

\begin{figure}[htbp]
\centering
\includegraphics[width=.8\linewidth]{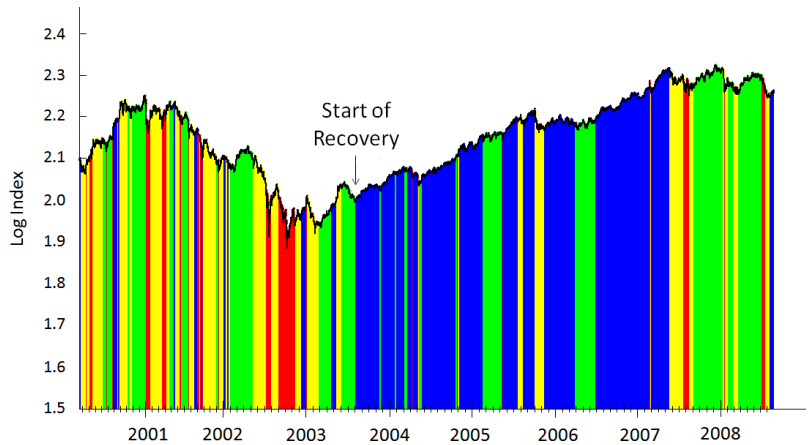}
\caption{Temporal distribution of the clustered segments for the time
series of UT, beween 14 February 2000 and 31 August 2008.  Based on
the working definition described in the text, the utilities sector
recovered from the mid-1998 to mid-2003 financial crisis on 6 August
2003 as indicated.}
\label{fig:startrecover}
\end{figure}

\begin{figure}[htbp]
\centering
\includegraphics[width=.8\linewidth]{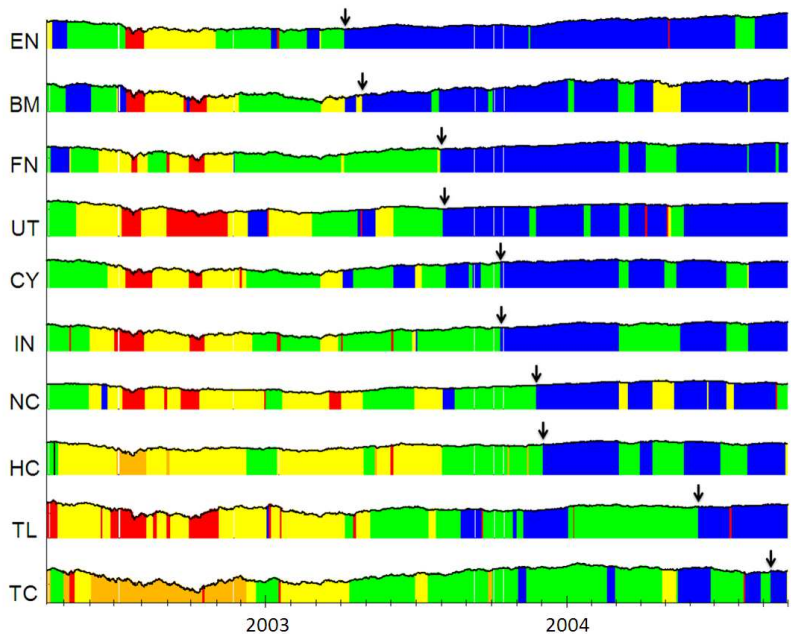}
\caption{Temporal distributions of clustered segments for the time
series of all ten US economic sectors between April 2002 and September
2004, showing the sequence of recovery from the mid-1998 to mid-2003
financial crisis.}
\label{fig:recoverysequence}
\end{figure}

As a working definition, an economic sector is deemed to have
recovered from the high-volatility phase, when we can identify in its
time series low-volatility segments that run for longer than two
months.  The choice of the two-month duration is arbitrary, but as
shown in the clustered segments of UT in Fig.~\ref{fig:startrecover}
for example, we find that the `post-recovery' time series always
consists predominantly of low-volatility segments, interrupted
infrequently by moderate-volatility market correction phases.  With
this definition, we find a very clear pattern of a temporally extended
recovery in the ten economic sectors from the previous financial
crisis.  As shown in Fig.~\ref{fig:recoverysequence}, EN and BM led
the US economic recovery around April/May 2003, followed by FN and UT
in early August 2003, CY and IN in mid-October 2003, NC and HC around
November/December 2004, TL in mid-June 2004, and finally TC in
mid-September 2004.  The time interval between the first sector
recovering and the last sector recovering is roughly one and a half
years.  Unless the present inner workings of the US economy is
entirely different from what it was ten years ago, we believe this is
the time scale US policy makers have to wrestle with to achieve
complete economic recovery from the current financial crisis.

From Fig.~\ref{fig:recoverysequence}, we see that the last sector to
recover from the previous financial crisis (mid-1998 to mid-2003) is
TC.  This is understandable, because the previous financial crisis was
the result of the technology bubble bursting, so it would be natural
for investors to stay away from the technology sector while the
economy is recovering.  The observation that TL is the second-to-last
sector to recover is also understandable: the fortunes of the
telecommunications sector is most strongly tied to that of the
technology sector.  In general, the more `basic' economic sectors
recover ahead of the more `advanced' economic sectors, because the
output of the former must be consumed by the latter to drive the
economic recovery.  The other feature clearly visible in
Fig.~\ref{fig:recoverysequence} is the pairwise recovery by (EN, BM),
(FN, UT), (CY, IN), and (NC, HC).  We suspect such pairings suggest
closer causal relationship between members of the pairs.  To ensure
that the pairings observed in the recovery sequence, and also in the
onset sequence we report in the next subsection, are economically
meaningful and not merely accidental, we will analyze this causal
proximity more carefully in Section \ref{sect:shockbyshock}.  An
understanding of the distribution of causal distances between economic
sectors is clearly critical to policy making.

\begin{figure}[htbp]
\centering\footnotesize
\begin{minipage}[t]{.475\linewidth}
\centering
\includegraphics[width=\linewidth]{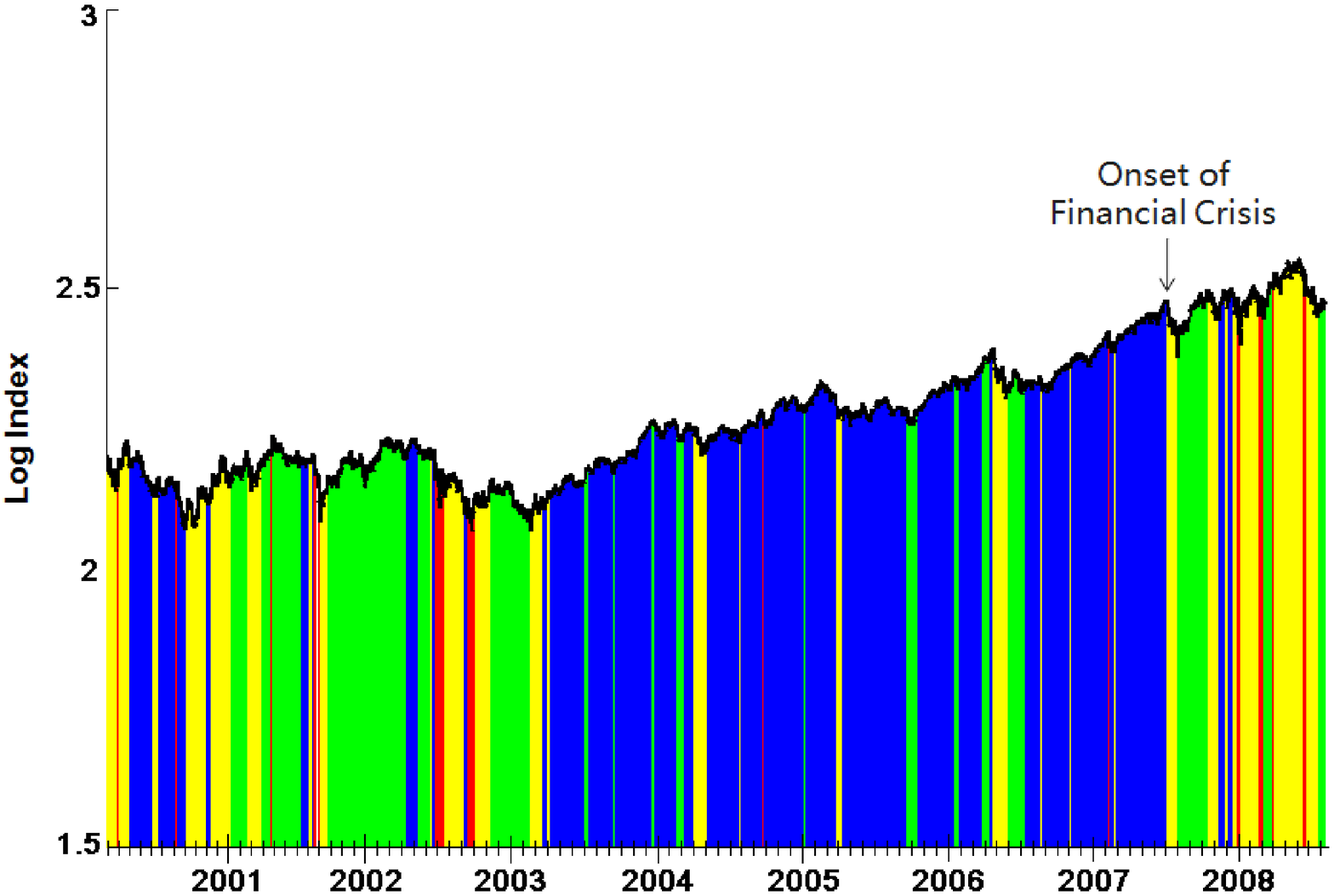}
\vskip .5\baselineskip
(a)
\end{minipage}
\hfill
\begin{minipage}[t]{.475\linewidth}
\centering
\includegraphics[width=\linewidth]{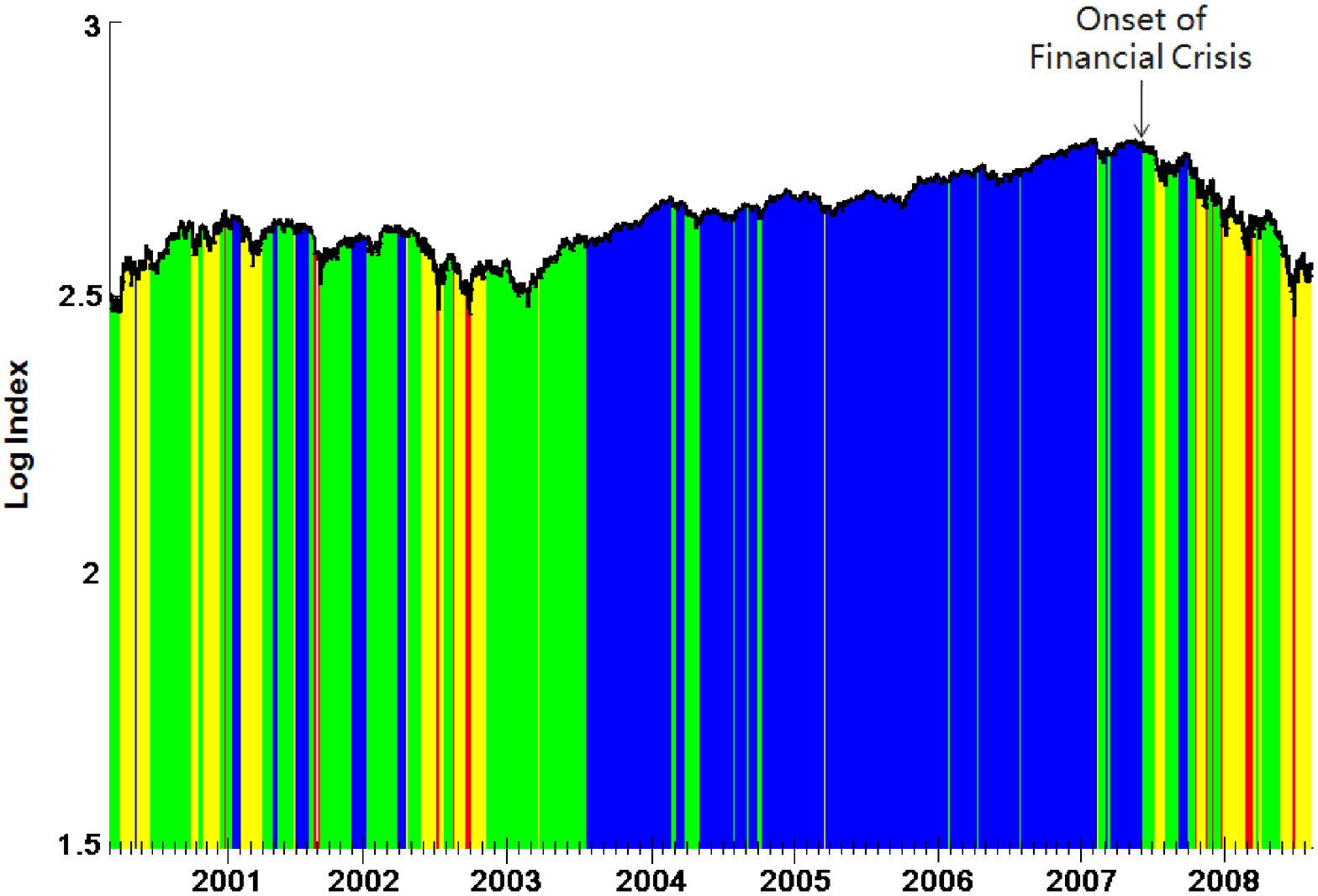}
\vskip .5\baselineskip
(b)
\end{minipage}

\begin{minipage}[t]{.475\linewidth}
\centering
\includegraphics[width=\linewidth]{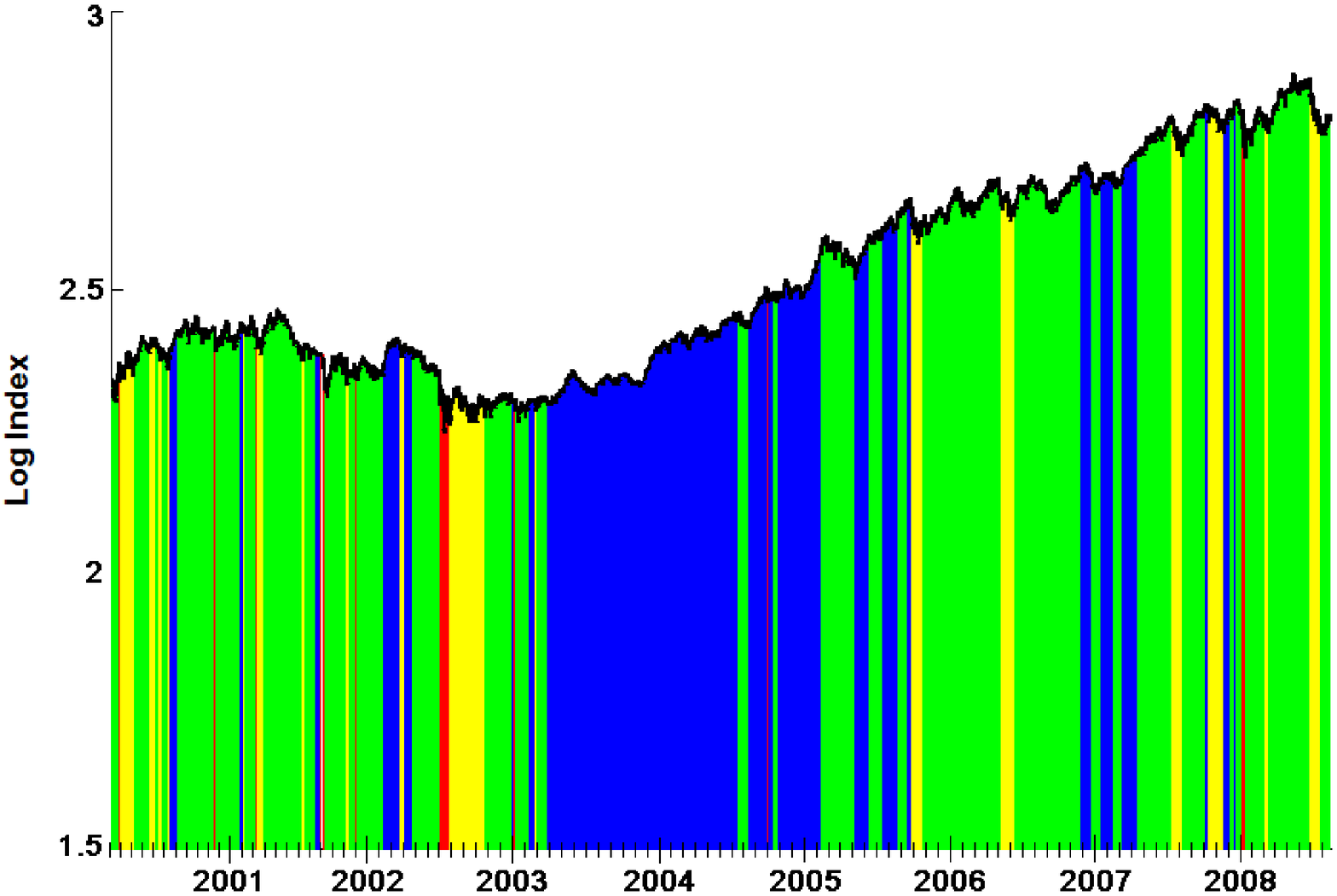}
\vskip .5\baselineskip
(c)
\end{minipage}
\hfill
\begin{minipage}[t]{.475\linewidth}
\centering
\includegraphics[width=\linewidth]{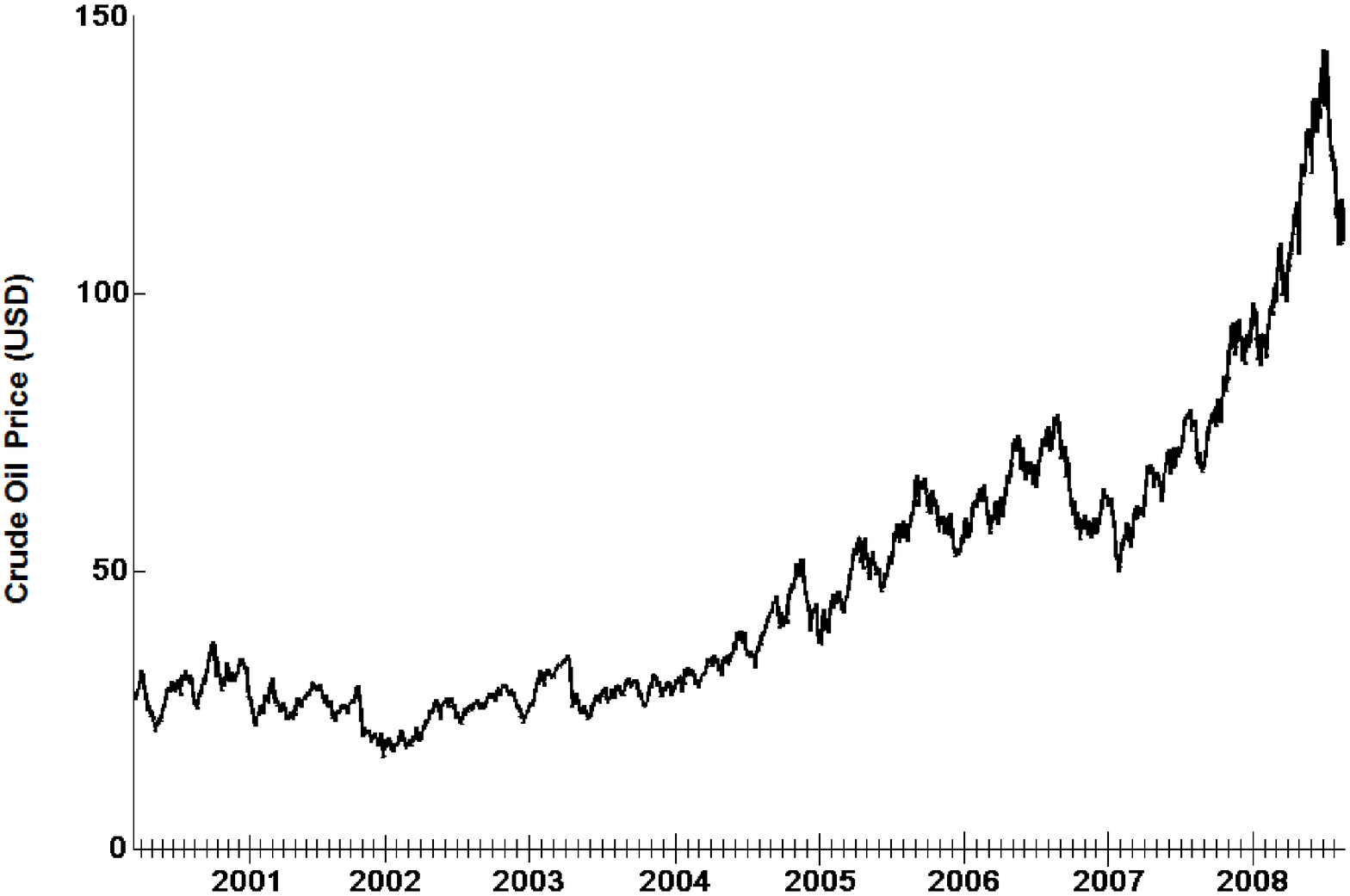}
\vskip .5\baselineskip
(d)
\end{minipage}
\caption{Temporal distributions of clustered segments between 14
February 2000 and 31 August 2008, showing the onsets of the present
financial crisis in the (a) BM and (b) FN sectors, and also the
anomaly in the (c) EN sector.  Also shown is the (d) price of crude
oil, which started rising sharply after the mid-2003 economic
recovery.}
\label{fig:startpresentcrisis}
\end{figure}

\subsection{Mid-2007 onset of current global financial crisis}

Analogous to our working definition of an economic recovery, we define
the start of the high-volatility phase (the present financial crisis)
in an economic sector time series as the end of the final
low-volatility phase lasting longer than two months.  In
Fig.~\ref{fig:startpresentcrisis}, we show the temporal distributions
of clustered segments for (a) BM, (b) FN, and (c) EN.  The start dates
of high-volatility phases in FN and BM are 20 June 2007 and 23 July
2007 respectively, consistent with our earlier finding that the
current global financial crisis started in July 2007.  The EN sector,
however, is an anomaly, because based on our working definition, the
start of the high-volatility phase for EN would be 24 February 2005.
In reality, the volatility of the EN sector time series is only
moderate between 2005 and 2007, so what we are seeing in
Fig.~\ref{fig:startpresentcrisis}(c) is an extremely extended market
correction phase, driven by the ever rising oil price
(Fig.~\ref{fig:startpresentcrisis}(d)).

\begin{figure}[htbp]
\centering
\includegraphics[width=.8\linewidth]{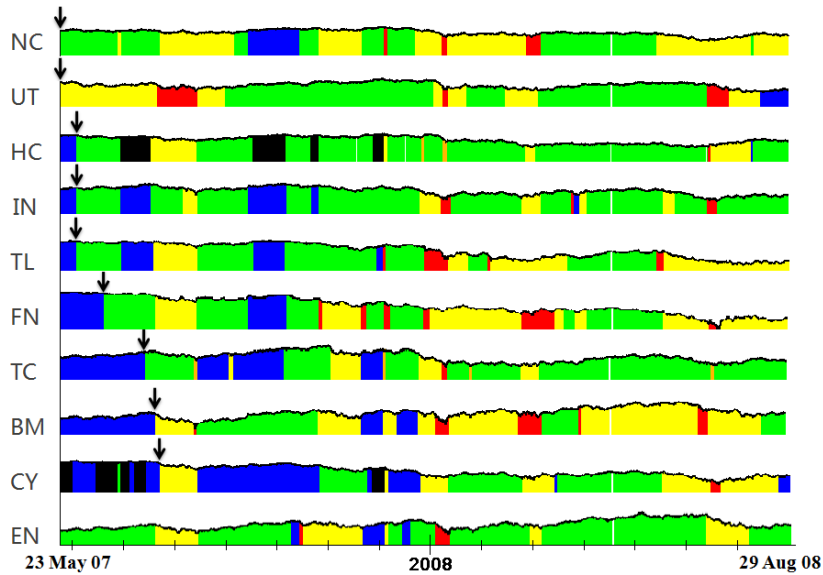}
\caption{Temporal distributions of clustered segments for the time
series of all ten US economic sectors between 23 May 2007 and 29
August 2008, showing the sequence of descent into the present
financial crisis.}
\label{fig:onsetsequence}
\end{figure}

This anomaly aside, the pattern of clustered segments observed (see
Fig.~\ref{fig:onsetsequence}) for the mid-2007 onset of the present
global financial crisis stands in stark contrast to the mid-2003
recovery from the previous financial crisis.  Firstly, the time scale
of the onset is very much shorter.  Starting with NC/UT on 23 May
2007, HC/IN/TL on 4 June 2007, FN on 20 June 2007, TC on 17 July 2007,
BM on 23 July 2007, and CY on 25 July 2007, the US economy went from
the first sector to the last sector into the high-volatility phase in
a mere two months!  Secondly, the synchronized or nearly-synchronized
groups of economic sectors are different: (NC, UT), (HC, IN, TL), (BM,
CY).  Thirdly, and most importantly, whereas the time series dynamics
during the mid-2003 economic recovery appears to be driven by
endogeneous interactions between the ten economic sectors, the time
series dynamics during the mid-2007 breakdown of the global financial
machinery appears to be driven by exogeneous factors.  We will discuss
in greater details this last observation in Section
\ref{sect:interestratecuts}.

To date, the many accounts \cite{Frank2008IMFWP08200,
Green2008JHousingEcon17p262, Lohr2008NYT, Lo2008Testimony,
Lo2009JFinEconPolicy1p1757, Neven2009ErasmusUnivMFEThesis,
Tudor2009RomEconJ31p115} of the present global financial crisis paint
a complex picture of how the crisis came to pass.  Our analysis
suggests that the reasons behind the collapse of investor confidence
worldwide might be even more complicated.  Instead of being led by a
crisis in the homebuilding and property industries, and the ensuing
waves of mortgage defaults silently catching up to the financial
institutions, we find the declines in the UT, HC, IN, and TL sectors
between those of NC and FN, whose downfalls were nearly one month
apart.  If we were to assume that these four sectors were not at
fault, and were merely collateral damage and early sacrifices of the
subprime excesses, then fully half of the US economic sectors were in
trouble before the financials found themselves in thick soup.  One
might wonder why no one saw and acted on these writings on the wall.

\section{Shock-by-shock causal-link analysis}
\label{sect:shockbyshock}

Perhaps no one understood these signs of the times, because they are
written in the language of statistical fluctuations.  In general,
people understand things better if they are cast in relational terms,
for example, cause and effect, leader and follower, \emph{etc}.  In
this section, we map out probable causal links between the ten US
economic sectors, based on the temporal distributions of the clustered
segments, and their associated statistics.  We do this first for the
entire high-volatility phase prior to the mid-2003 economic recovery,
and then for corresponding high-volatility shocks (to be defined
later) preceding the mid-2003 economic recovery.  Finally, we analyze
corresponding high-volatility shocks after the July 2007 onset of the
present financial crisis.

\subsection{The entire high-volatility phase prior to mid-2003
economic recovery}

\begin{figure}[htbp]
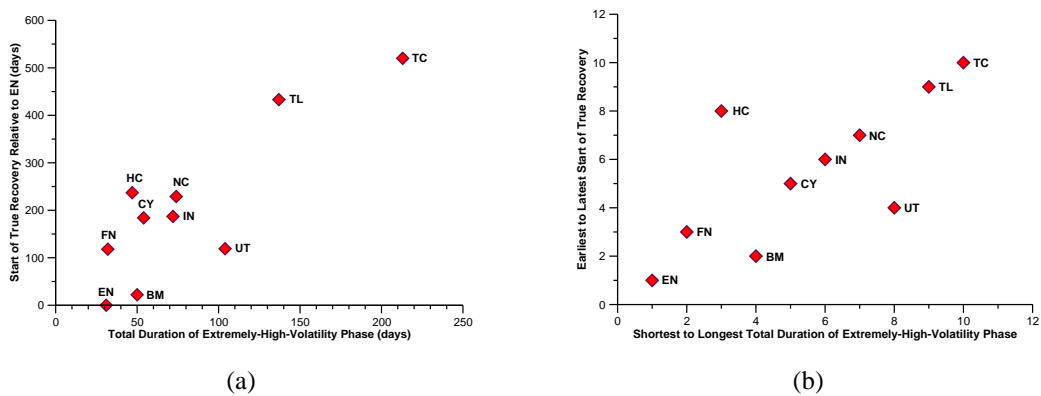

\centering\footnotesize
\begin{minipage}[t]{.45\linewidth}
\centering
\includegraphics[width=\linewidth,clip=true]{scatterstartvsduration}
\vskip .5\baselineskip
(a)
\end{minipage}
\hfill
\begin{minipage}[t]{.45\linewidth}
\centering
\includegraphics[width=\linewidth,clip=true]{rankstartvsduration}
\vskip .5\baselineskip
(b)
\end{minipage}
\caption{(a) Scatter plot and (b) rank plot of the start of economic
recovery against duration of the high-volatility phase preceding the
mid-2003 recovery from the previous financial crisis.  As expected
from causal tree analogy, there is a positive correlation between when
the economic recovery starts, and how long the high-volatility phase
lasts.  Surprising outliers such as HC and UT can also be clearly
identified from the rank plot.}
\label{fig:startvsduration}
\end{figure}

From Fig.~\ref{fig:recoverysequence}, we can very roughly see that the
later the start of recovery, the longer the high-volatility phase.
This positive correlation between the start of recovery and duration
of high-volatility phase can be seen more clearly in the form of a
scatter plot (Fig.~\ref{fig:startvsduration}(a)).  However, causal
relationships between economic sectors are not so clear from the
scatter plot.  When the same information is presented as a rank plot
in Fig.~\ref{fig:startvsduration}(b), the causal relationships become
clearer.  In fact, only in the rank plot is it clear that HC and UT
are outliers during the recovery from the previous financial crisis.
In going from a scatter plot to a rank plot, we have gone from
parametric statistics to nonparametric order statistics.  It is well
known in the statistics community that order statistics, being much
less sensitive to how well parameters are estimated, allows us to
arrive at much more robust conclusions on trends and outliers
\cite{DavidNagaraja2003, ArnoldBalakrishnanNagaraja}.

It would be interesting to perform a similar analysis for the present
financial crisis, to compare and to contrast.  However, at the time of
writing, this crisis is not yet over.  It is also tempting, based on
the robustness of the rank plot results shown in
Fig.~\ref{fig:startvsduration}(b), to expect EN, FN and BM to again
lead the recovery (or at least be very close to the start of the
complete recovery), and that recovery in CY will precede IN, which
will in turn precede NC, as suggested by
Fig.~\ref{fig:startvsduration}(b).  However, we must remind ourselves
of the difference between robustness and significance.  A given
sequence is only meaningful, and predictive, only if it is robustly
determined, and its appearance statistically significant compared
against the null hypothesis of the sequence appearing by chance.  To
establish that temporal proximity in their response is a statistical
significant indicator of causal proximity between two US economic
sectors, we analyze various rank plots of the ten US economic sectors
at the shock-by-shock level, for both the period prior to the mid-2003
economic recovery, as well as the period after the mid-2007 start of
the Subprime Crisis.  Basically, a significant and robust sequence is
one that we expect to see over and over again in the following
shock-by-shock analysis.

\subsection{Very-high-volatility shocks prior to mid-2003 economic
recovery}

\begin{figure}[htbp]
\centering
\includegraphics[width=.8\linewidth]{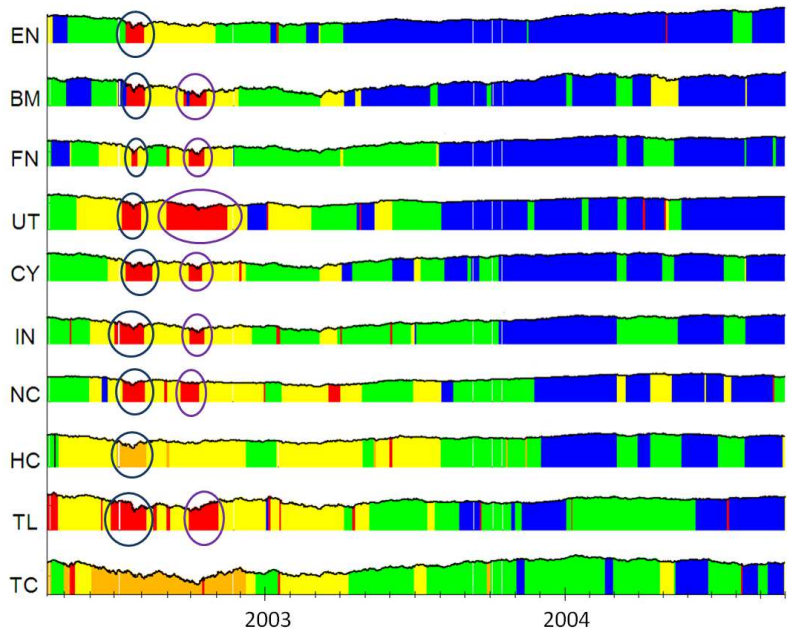}
\caption{The pair of very-high-volatility shocks identified in most US
economic sectors.  The July 2002 shock is missing from HC and TC, but
can be substituted for by a high-volatility shock in HC.  The same
cannot be done for TC, since its high-volatility shock straddles both
the July 2002 and October 2002 shocks.  The October 2002 shock is
missing from EN, HC, and TC.  No substitutions for these missing
shocks can be made.}
\label{fig:sigredshocks}
\end{figure}

To repeat the causal-link analysis presented above on the more
detailed level of individual shocks, which are periods in the time
series characterized by high market volatilities, we observe the
presence of \emph{corresponding shocks} in the different sectors.
These can be identified based on the volatility, or the direction of
volatility change.  In Fig.~\ref{fig:sigredshocks}, we highlight a
pair of very-high-volatility shocks (July 2002 and October 2002)
experienced by most of the economic sectors around the 2002 low of the
major US indices.  For most economic sectors, this is the last
extended very-high-volatility shocks experienced prior to the mid-2003
economic recovery.

\begin{figure}[htbp]
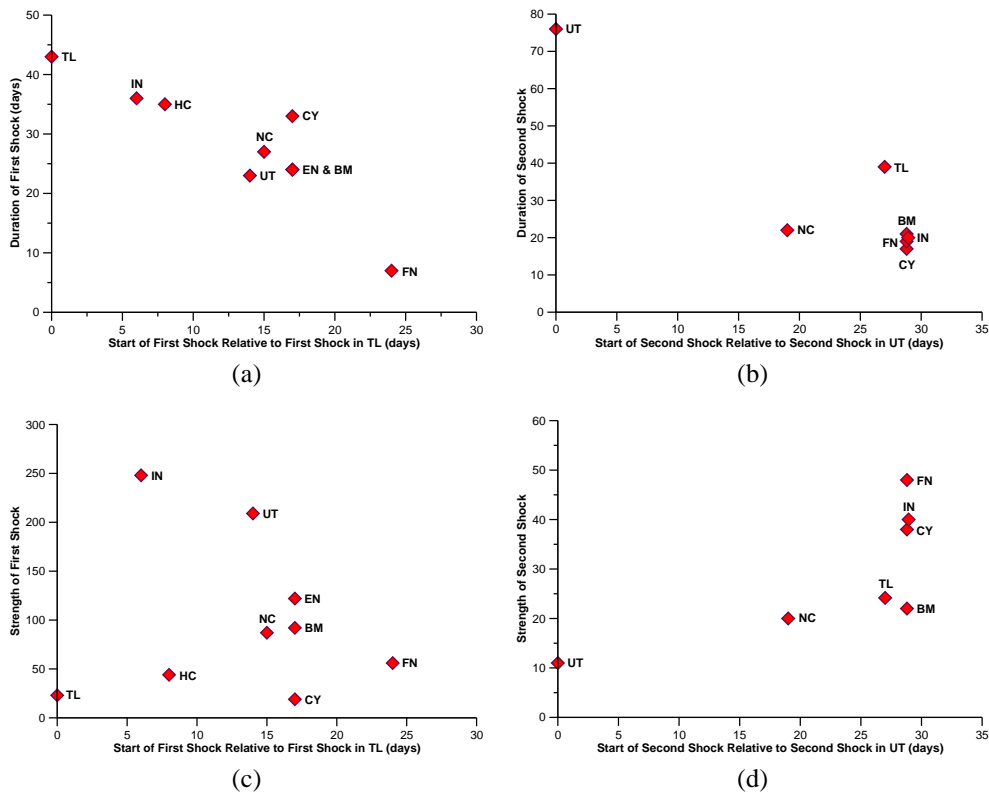

\centering\footnotesize
\begin{tabular}{cc}
\includegraphics[width=.46\linewidth,clip=true]{red1scatterdurationvsstart} &
\includegraphics[width=.46\linewidth,clip=true]{red2scatterdurationvsstart} \\
(a) & (b) \\ [2ex]
\includegraphics[width=.46\linewidth,clip=true]{red1scatterJSvsstart} &
\includegraphics[width=.46\linewidth,clip=true]{red2scatterJSvsstart} \\
(c) & (d) \\
\end{tabular}
\caption{Scatter plots of duration against starting time (top, (a) for
July 2002 shock, and (b) for October 2002 shock) and of strength (as
measured by the Jensen-Shannon divergence of the leading boundary of
the shock) against starting time (bottom, (c) for July 2002 shock, and
(d) for October 2002 shock).}
\label{fig:red12scatter}
\end{figure}

In Fig.~\ref{fig:red12scatter} we show the scatter plots of the
duration and strength of the shocks against the start of the shocks.
As expected from our causal tree analogy, there is a negative
correlation between duration of shock and start of shock.  This
relationship is clearer for the first (July 2002) shock
(Fig.~\ref{fig:red12scatter}(a)), and less clear for the second
(October 2002) shock (Fig.~\ref{fig:red12scatter}(b)), where the
starting dates of BM, CY, FN, IN and TL are highly clustered.  Based
on our causal tree analogy, we also expect a leading sector to
experience a stronger shock compared to a trailing sector.  This
translates to an expected negative correlation between the
Jensen-Shannon divergence value of the leading boundary of the
very-high-volatility shocks and the start of the shock.  Here, let us
recall that the larger $\Delta^*$ is, the more statistically different
the very-high-volatility shock is from the preceding segment, and thus
the stronger the shock. 

\begin{figure}[htb]
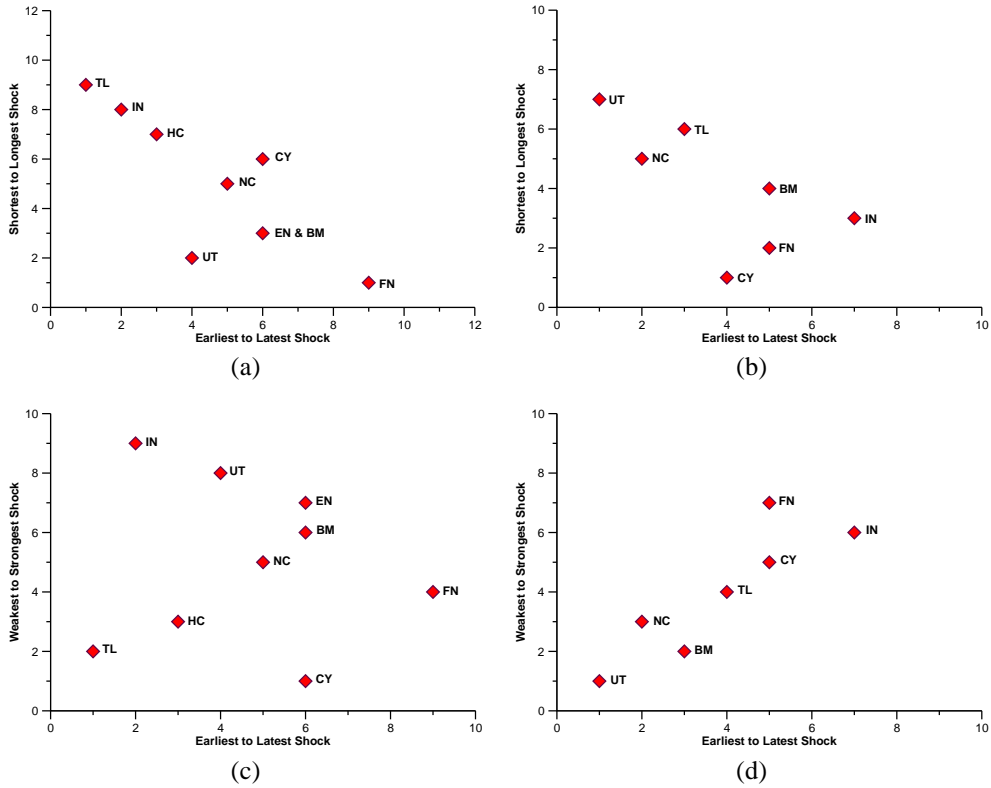

\centering\footnotesize
\begin{tabular}{cc}
\includegraphics[width=.46\linewidth,clip=true]{red1rankdurationvsstart} &
\includegraphics[width=.46\linewidth,clip=true]{red2rankdurationvsstart} \\
(a) & (b) \\ [2ex]
\includegraphics[width=.46\linewidth,clip=true]{red1rankJSvsstart} &
\includegraphics[width=.46\linewidth,clip=true]{red2rankJSvsstart} \\
(c) & (d) \\
\end{tabular}
\caption{Rank plots of duration against starting time (top, (a) for
July 2002 shock, and (b) for October 2002 shock) and of strength (as
measured by the Jensen-Shannon divergence of the leading boundary of
the shock) against starting time (bottom, (c) for July 2002 shock, and
(d) for October 2002 shock).}
\label{fig:red12rank}
\end{figure}

However, the scatter plots (c) and (d) in Fig.~\ref{fig:red12scatter}
suggests something more complex and more interesting.  For the first
shock, a negative correlation between Jensen-Shannon divergence and
start was observed for IN, UT and FN, but a positive correlation is
observed for TL, HC, NC, BM, and EN.  In this first shock, CY is an
outlier, which is not clear from Fig.~\ref{fig:red12scatter}(a).  For
the second shock, we see surprisingly from
Fig.~\ref{fig:red12scatter}(d) a strictly positive trend for all
participating sectors.  These trends, which can be seen even clearer
on the rank plots shown in Fig.~\ref{fig:red12rank}, can only be
explained if there is amplification as volatility shocks propagate
from one sector to another in the causal tree.  If this amplification
is linear, and is weaker than the expected linear dissipation, we
would expect to see dissipative propagation of volatility shocks all
the time.  Conversely, if the linear amplification is stronger than
the linear dissipation, we would expect to see amplified propagation
of volatility shocks all the time.  In our plots, we see mostly
dissipative propagation of volatility shocks, and amplification only
occasionally.  This suggests that the amplification of volatility
shocks is of a nonlinear nature.

\subsection{High-volatility shocks after mid-2007 onset of present
financial crisis}

\begin{figure}[htbp]
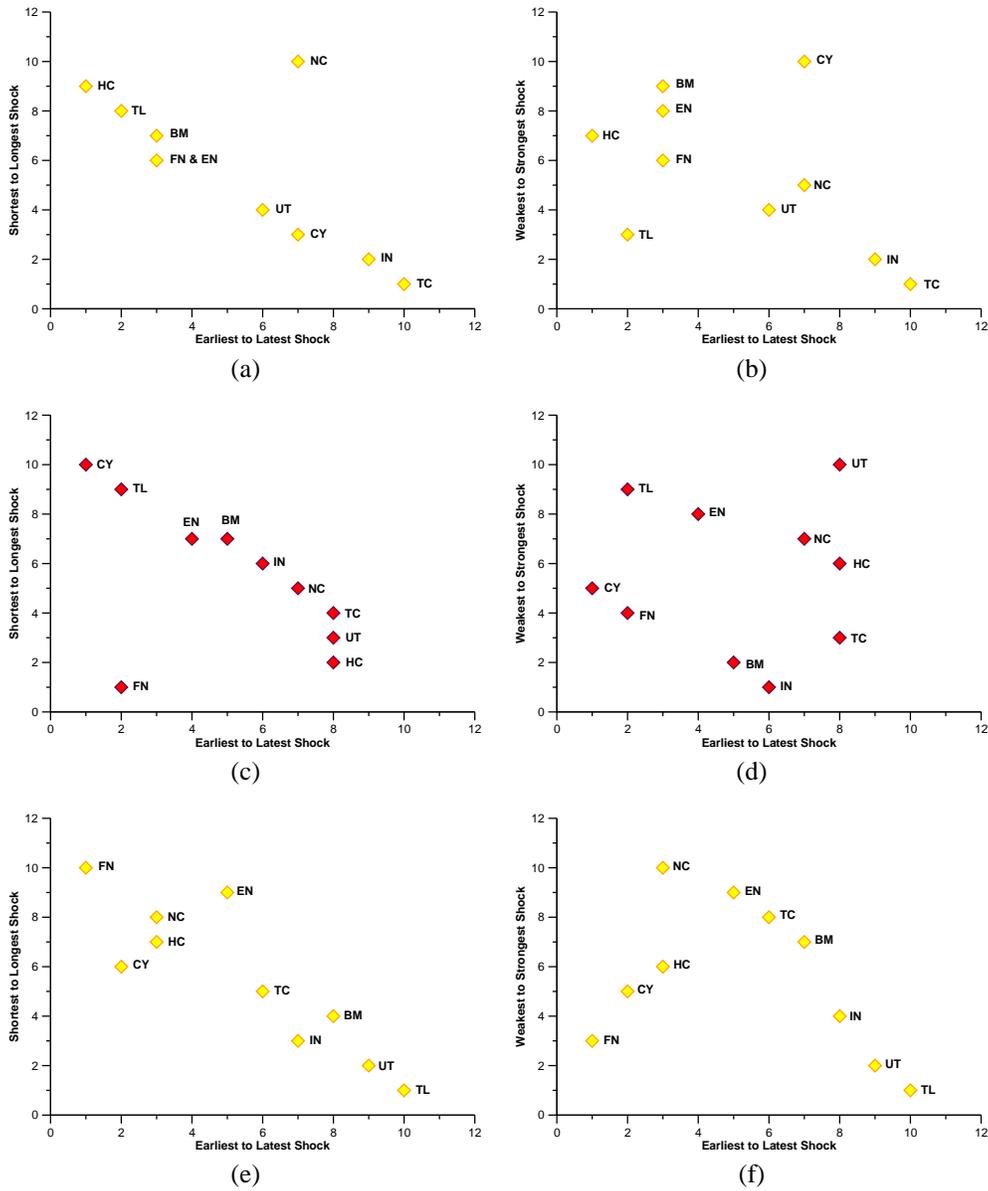

\centering\footnotesize
\begin{tabular}{cc}
\includegraphics[width=.46\linewidth,clip=true]{yellow1rankdurationstart} &
\includegraphics[width=.46\linewidth,clip=true]{yellow1rankJSstart} \\
(a) & (b) \\ [2ex]
\includegraphics[width=.46\linewidth,clip=true]{red1rankdurationstart} &
\includegraphics[width=.46\linewidth,clip=true]{red1rankJSstart} \\
(c) & (d) \\ [2ex]
\includegraphics[width=.46\linewidth,clip=true]{yellow2rankdurationstart} &
\includegraphics[width=.46\linewidth,clip=true]{yellow2rankJSstart} \\
(e) & (f) \\
\end{tabular}
\caption{Rank plots of duration against starting time (left column)
and of strength against starting time (right column) for (i) the
end-July 2007 high-volatility shock ((a) and (b)), (ii) the
mid-January 2008 very-high-volatility shock ((c) and (d)), and (iii)
the end-January 2008 high-volatility shock ((e) and (f)) showing the
positive correlation expected from the causal tree analogy.  Apart
statistical outliers, the most prominent features seen in these plots
are the surprising organization of economic sectors into two clusters,
(CY, FN, BM, IN), and (TL, EN, NC, HC) during the mid-January 2008
very-high-volatility shock, and the unexpectedly weak high-volatility
shocks experienced by FN, CY, and HC.}
\label{fig:yellow12red1rank}
\end{figure}

As we shall show in the next section, the temporal distributions of
clustered segments right after the onset of the present financial
crisis appears to be strongly driven by the Federal Reserve interest
rate cuts.  The clustered segment boundaries between some economic
sectors coincide to within a day or two of the the dates interest
rates are revised.  This observation is highly significant, given the
statistical significance and robustness of the clustered segment
boundaries.  However, because of this strong driving, nearly all
scatter plots are highly clustered and not very informative.  In
contrast, rank plots constructed based on the precise half-hour of the
segment boundaries are more informative.  In
Fig.~\ref{fig:yellow12red1rank}, we show the rank plots for two
high-volatility shocks (end-July 2007 and end-January 2008) and one
very-high-volatility shock (mid-January 2008).

In Fig.~\ref{fig:yellow12red1rank}, we find the negative correlations
between duration and starting time, as well as between strength and
starting time for all three shocks expected from the causal tree
analogy.  However, there are several surprises.  First, different
statistical outliers are identified from different shocks, and also
from different rank plots.  For the first high-volatility shock
(end-July 2007), NC experienced a much longer shock, whereas CY
experienced a much stronger shock than expected from their respective
starting times.  For the very-high-volatility shock (mid-January
2008), FN experienced a much shorter shock, whereas UT experienced a
stronger shock than expected from their respective starting times.
For the second high-volatility shock (end-January 2008) immediately
following the very-high-volatility shock, we find FN, CY, and HC
experiencing much weaker shocks than expected based on the starting
times of their respective shocks.  As suggested earlier, this is the
signature of nonlinear amplification of the volatility shock as it
propagates from FN to CY to HC.  Second, from the rank plot of
strength against starting time of the very-high-volatility shock, we
detect two different clusters of economic sectors, (CY, FN, BM, IN)
and (TL, EN, NC, HC).  This statistical feature suggests that there
are two (or perhaps more) shocks propagating in tandem through the US
economy around mid-January 2008.  Third, the orderings in the duration
versus starting time rank plots are different for the first and second
high-volatility shocks, suggesting that causal links in the US economy
are highly dynamic, and constantly rearranging themselves.

\begin{figure}[htb]
\centering
\includegraphics[width=.8\linewidth]{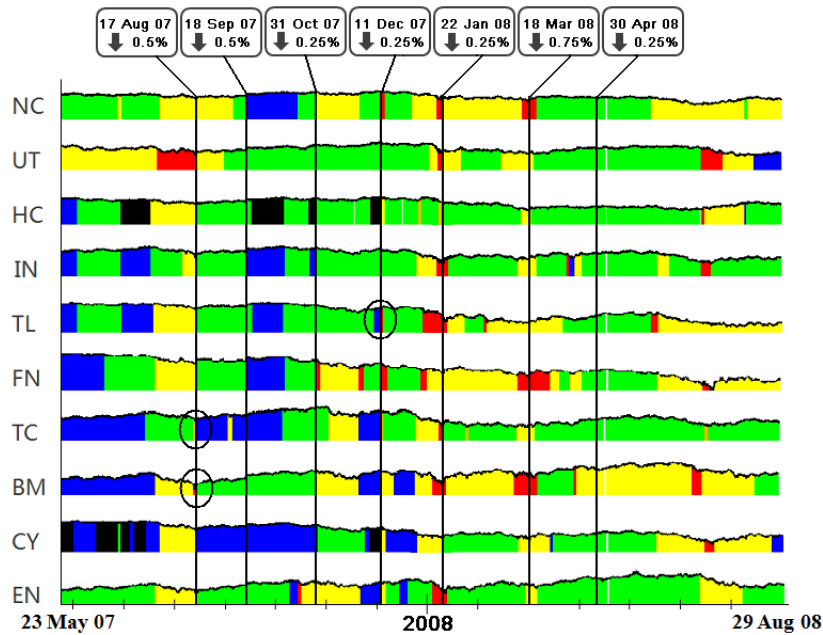}
\caption{Federal reserve interest rate cuts during the onset of the
present financial crisis, superimposed onto the temporal distributions
of clustered segments of the ten US economic sectors between 23 May
2007 and 29 August 2008.  Assuming the goal of an interest rate cut is
to calm the market down, we find the first two cuts effective, the
next three cuts counter-effective, and the last two cuts ineffective.
Also shown are circles indicating anticipation of the interest rate
cuts on 17 August 2007 and 11 December 2007.}
\label{fig:onsetinterestratecuts}
\end{figure}

\section{The (near) futility of interest rate cuts}
\label{sect:interestratecuts}

After the subtle and statistically weaker onset signatures between May
and July 2007, most of the important shocks in different economic
sectors occur within a day or two of each other, and appear to be
exogeneously driven by Federal Reserve interest rate cuts, as shown in
Fig.~\ref{fig:onsetinterestratecuts}.  In BM, TC, and TL, we see brief
volatility movements a few days to a week before two interest rate
cuts (circled in Fig.~\ref{fig:onsetinterestratecuts}), suggesting
that these sectors were anticipating the rate cuts.  Naturally, all fiscal
policies are double-edged swords.  According to Investopedia
\cite{investopedia}, a decrease in interest rates is supposed to move
money from the bond market to the stock market.  It is also supposed
to allow businesses to finance their expansion at a cheaper rate,
increase their future earnings, and thereby bring about higher stock
prices.  Even though an interest rate cut erodes the banks' ability
to make money (since the main business of banks is to lend money), the
overall psychological impact of interest rate cuts is regarded as
positive, more so during a financial crisis.  Taking the investors'
sentiments into account, it is likely that interest rate cuts
were implemented during the onset of the Subprime Crisis to calm the
market.

Indeed, looking at Fig.~\ref{fig:onsetinterestratecuts}, the first
rate cut on 17 August 2007 appears to be highly effective, in the
sense that market volatilities fell across a broad spectrum of
economic sectors right after the cut.  The only exception is NC, which
did not respond to this first rate cut.  In comparison, the second
rate cut on 18 September 2007 appears to be slightly less effective.
On 18 September, TC and CY were already in the low-volatility phase,
so we do not expect the second rate cut to do anything to these
sectors anyway.  However, even after factoring in anticipations and
lags, BM and EN, which were in the moderate-volatility phase, did not
respond to this second rate cut.  More interestingly, the next three
rate cuts, on 31 October 2007, 11 December 2007, and 22 January 2008,
appear to have the opposite effect as intended, increasing (instead of
lowering) market volatilities in a number of economic sectors.  Most
notably, NC, HC, TC and BM reacted adversely to all three rate cuts.
Finally, we observe that the last two rate cuts, on 18 March 2008 and
30 April 2008, do not coincide with any of the clustered segment
boundaries.  As far as we can tell, the interest rate cuts were
ineffective in evoking any kind of response from the market.  This is
especially true for the last rate cut on 30 April 2008.

\begin{figure}[htbp]
\centering\footnotesize
\includegraphics[width=0.7\linewidth]{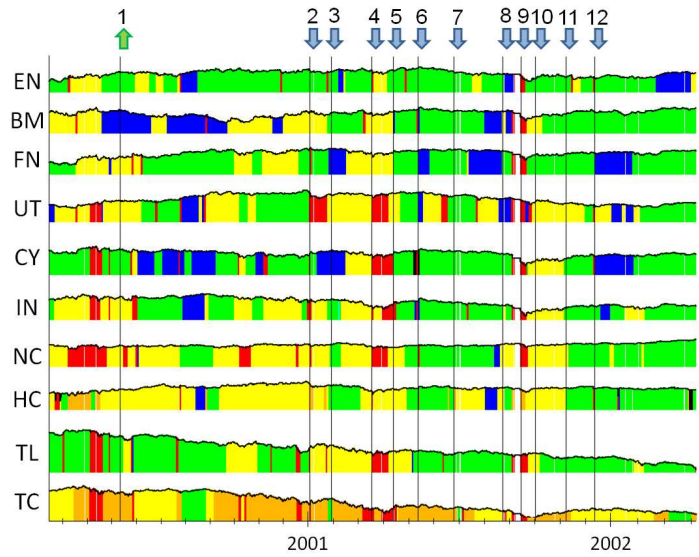}

(a)

\includegraphics[width=0.7\linewidth]{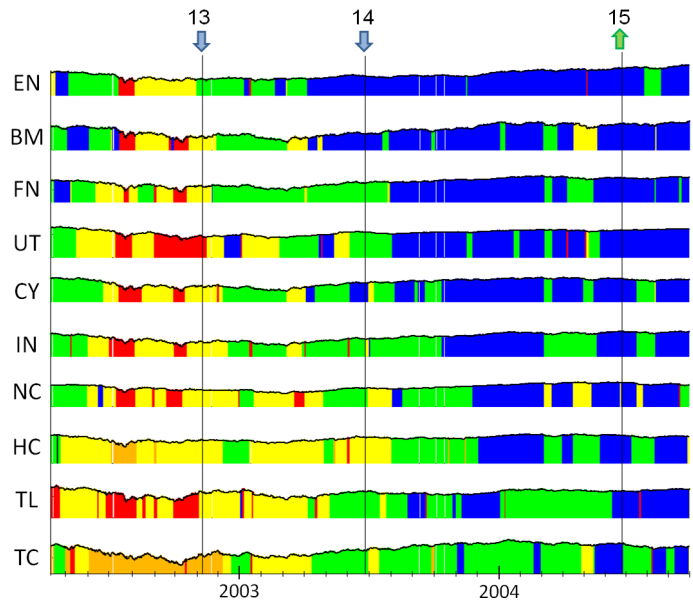}

(b)
\caption{Federal Reserve interest rate adjustments made (a) between
March 2000 and March 2002, prior to the 2002 low for major US indices,
and (b) between April 2002 and September 2004, superimposed onto the
temporal distributions of clustered segments of the ten US economic
sectors.  The interest rate cut alternates between being
counter-effective and effective, before becoming mostly ineffective.}
\label{fig:prerecoveryintratecuts}
\end{figure}

Similar interest rate cut driven dynamics were seen in 2001, the year
the US economy was officially in recession.  However, the picture on
whether interest rate cuts by the Federal Reserve (see Table
\ref{table:intratechange}) is an effective tool for macroeconomic
manipulations is a lot less clear.  As we can see from
Fig.~\ref{fig:prerecoveryintratecuts}(a), successive interest rate
cuts alternates between counter-effective (market volatilities
increased in most economic sectors) and effective (market volatilities
decreased in most economic sectors), before losing its effectiveness
(no change in market volatilities for most economic sectors) in the
last few cuts.  This ineffectiveness of interest rate cuts continued
into 2002 and 2003 (see Fig.~\ref{fig:prerecoveryintratecuts}(b)),
when the US economy started recovering from the previous financial
crisis.

\begin{table}[htbp]
\centering\footnotesize
\caption{Changes in the Federal Reserve interest rate between February
2000 and September 2004.}
\label{table:intratechange}
\vskip .5\baselineskip
\begin{tabular}{cccc}
\hline
S/No. & Date & Change (\%) & New Rate (\%) \\
\hline
1 & 16 May 2000 & $+0.50$ & 6.50 \\
2 & 3 January 2001 & $-0.50$ & 6.00 \\
3 & 31 January 2001 & $-0.50$ & 5.50 \\
4 & 20 March 2001 & $-0.50$ & 5.00 \\
5 & 18 April 2001 & $-0.50$ & 4.50 \\
6 & 15 May 2001 & $-0.50$ & 4.00 \\
7 & 27 June 2001 & $-0.25$ & 3.75 \\
8 & 21 August 2001 & $-0.25$ & 3.50 \\
9 & 17 September 2001 & $-0.50$ & 3.00 \\
10 & 2 October 2001 & $-0.50$ & 2.50 \\
11 & 6 November 2001 & $-0.50$ & 2.00 \\
12 & 11 December 2001 & $-0.25$ & 1.75 \\
13 & 6 November 2002 & $-0.50$ & 1.25 \\
14 & 25 June 2003 & $-0.25$ & 1.00 \\
15 & 30 June 2004 & $+0.25$ & 1.25 \\
\hline
\end{tabular}
\end{table}

\section{Conclusions}
\label{sect:conclusions}

To understand the causal links and processes within the US economy, we
performed in this paper a comparative segmentation and clustering
analysis of the time series data between 14 February 2000 and 31
August 2008 for the ten Dow Jones US economic sector indices.  Based
on general features of the temporal distributions of clustered
segments, we see a clear pattern of economic recovery from the
mid-1998 to mid-2003 financial crisis, and also a clear pattern of
descent into the present global financial crisis.  In particular, we
saw how EN and BM led the one-and-a-half-year long recovery from the
previous financial crisis precipitated by TC, and how NC and UT led the
two-month-long decline into the present financial crisis.  Apart from
the greatly differing time scales between recovery and onset, our
study also reveals that on a macroscopic scale, the economic sector
going first into a financial crisis recovers the last, and the last
economic sector to be in trouble recovers first.

From the temporal distributions of clustered segments, we were also
able to identify corresponding shocks in the different economic
sectors, based on the volatility, or the direction of volatility
change.  Our shock-by-shock causal-link analysis thereafter reveals on
a mesoscopic level that leading sectors experience a stronger and
longer shock, whereas trailing sectors experience a weaker and shorter
shock.  We also observe in general that corresponding shocks start in
close temporal proximity to each other within the most closely related
sectors.  These general observations are robust, because they are
derived from the order statistics of the starting dates, durations,
and strengths of shocks in the various economic sectors.  These
observations are also modestly significant, even though the sample
size is small, because they are repeatedly observed for different
shocks in different historical periods.  More importantly, these
general observations are consistent with the causal tree analogy we
developed, which helps us simplify our mental model of the response of
an economy to financial crises.  In addition to dissipative
propagation of volatility shocks from one economic sector to another,
we also find evidences for nonlinear amplification, and complex
sectorial structures for the propagating shocks that suggest a highly
dynamic US economy.

Most interestingly, while the mid-2003 economic recovery appears to be
driven by endogeneous interactions within the US economy, the dynamics
during the mid-2007 onset of the Subprime Crisis appears to be
strongly driven by the Federal Reserve interest rate cuts.  By
comparing the dates interest rates were cut to the statistical
significant boundaries of clustered segments, we find that the first
few interest rate cuts are effective, i.e. market volatility decreases
across a broad spectrum of economic sectors.  Surprisingly, the next
few interest rate cuts are counter-effective, in the sense that the
market volatility increases after the rate cut in most economic
sectors.  Thereafter, the volatilities in most economic sectors stop
responding to further interest rate cuts, which have thus become
ineffective.  A slightly more complex pattern of interest rate cuts
alternating between effective and counter-effective, before becoming
ineffective, was also found during 2001 (the year the US economy
officially went into a recession).  The moral of the story is clear:
an interest rate cut by the Federal Reserve is not a magic bullet, nor
panacea for all our economic woes, but must be administered sparingly
to be effective.

\section*{Acknowledgements}

This research is supported by the Nanyang Technological University
startup grant SUG 19/07.  We have had helpful discussions with Chris
Kok Jun Liang and Tan Chong Hui.

\appendix

\section{Top components of Dow Jones US economic sector indices}
\label{app:components}

\subsection{Basic Materials}

\begin{center}\footnotesize
\begin{tabular}{clc}
\hline
ISIN/Ticker & \parbox[c]{7cm}{\centering Company} & Adjusted Weight \\
\hline
FCX & Freeport-McMoRan Copper \& Gold Inc. & 10.29\% \\
DD & E.I. DuPont de Nemours \& Co. & 9.12\% \\
DOW & Dow Chemical Co. & 7.70\% \\
NEM & Newmont Mining Corp. & 6.14\% \\
PX & Praxair Inc. & 6.13\% \\
APD & Air Products \& Chemicals Inc. & 3.67\% \\
BTU & Peabody Energy Corp. & 3.38\% \\
AA & Alcoa Inc. & 2.89\% \\
PPG & PPG Industries Inc. & 2.78\% \\
ECL & Ecolab Inc. & 2.41\% \\
\hline
\end{tabular}
\end{center}

\subsection{Consumer Services}

\begin{center}\footnotesize
\begin{tabular}{clc}
\hline
ISIN/Ticker & \parbox[c]{7cm}{\centering Company} & Adjusted Weight \\
\hline
WMT & Wal-Mart Stores Inc. & 7.28\% \\
MCD & McDonald's Corp. & 5.33\% \\
DIS & Walt Disney Co. & 4.15\% \\
AMZN & Amazon.com Inc. & 3.90\% \\
HD & Home Depot Inc. & 3.28\% \\
CVS & CVS Caremark Corp. & 2.70\% \\
CMCSA & Comcast Corp. Cl A & 2.64\% \\
TGT & Target Corp. & 2.44\% \\
DTV & DIRECTV Group Inc. & 2.30\% \\
WAG & Walgreen Co. & 2.18\% \\
\hline
\end{tabular}
\end{center}

\subsection{Oil \& Gas}

\begin{center}\footnotesize
\begin{tabular}{clc}
\hline
ISIN/Ticker & \parbox[c]{7cm}{\centering Company} & Adjusted Weight \\
\hline
XOM & Exxon Mobil Corp. & 25.57\% \\
CVX & Chevron Corp. & 11.68\% \\
SLB & Schlumberger Ltd. & 7.59\% \\
COP & ConocoPhillips & 6.01\% \\
OXY & Occidental Petroleum Corp. & 5.14\% \\
APA & Apache Corp. & 2.96\% \\
HAL & Halliburton Co. & 2.47\% \\
APC & Anadarko Petroleum Corp. & 2.28\% \\
DVN & Devon Energy Corp. & 2.08\% \\
NOV & National Oilwell Varco Inc. & 1.84\% \\
\hline
\end{tabular}
\end{center}

\subsection{Financials}

\begin{center}\footnotesize
\begin{tabular}{clc}
\hline
ISIN/Ticker & \parbox[c]{7cm}{\centering Company} & Adjusted Weight \\
\hline
JPM & JPMorgan Chase \& Co. & 7.50\% \\
WFC & Wells Fargo \& Co. & 6.73\% \\
BAC & Bank of America Corp. & 5.50\% \\
C & Citigroup Inc. & 5.04\% \\
GS & Goldman Sachs Group Inc. & 3.40\% \\
BRK/B & Berkshire Hathaway Inc. Cl B & 3.38\% \\
AXP & American Express Co. & 2.32\% \\
USB & U.S. Bancorp & 2.30\% \\
V & VISA Inc. Cl A & 1.85\% \\
BK & Bank of New York Mellon Corp. & 1.65\% \\
\hline
\end{tabular}
\end{center}

\subsection{Healthcare}

\begin{center}\footnotesize
\begin{tabular}{clc}
\hline
ISIN/Ticker & \parbox[c]{7cm}{\centering Company} & Adjusted Weight \\
\hline
JNJ & Johnson \& Johnson & 12.56\% \\
PFE & Pfizer Inc. & 9.68\% \\
MRK & Merck \& Co. Inc. & 7.82\% \\
ABT & Abbott Laboratories & 5.28\% \\
AMGN & Amgen Inc. & 3.72\% \\
BMY & Bristol-Myers Squibb Co. & 3.20\% \\
UNH & UnitedHealth Group Inc. & 3.03\% \\
MDT & Medtronic Inc. & 2.68\% \\
LLY & Eli Lilly \& Co. & 2.44\% \\
GILD & Gilead Sciences Inc. & 2.26\% \\
\hline
\end{tabular}
\end{center}

\subsection{Industrials}

\begin{center}\footnotesize
\begin{tabular}{clc}
\hline
ISIN/Ticker & \parbox[c]{7cm}{\centering Company} & Adjusted Weight \\
\hline
GE & General Electric Co. & 10.45\% \\
UTX & United Technologies Corp. & 4.03\% \\
MMM & 3M Co. & 3.39\% \\
UPS & United Parcel Service Inc. Cl B & 3.10\% \\
CAT & Caterpillar Inc. & 2.98\% \\
UNP & Union Pacific Corp. & 2.77\% \\
BA & Boeing Co. & 2.58\% \\
EMR & Emerson Electric Co. & 2.57\% \\
HON & Honeywell International Inc. & 2.15\% \\
DE & Deere \& Co. & 1.95\% \\
\hline
\end{tabular}
\end{center}

\subsection{Consumer Goods}

\begin{center}\footnotesize
\begin{tabular}{clc}
\hline
ISIN/Ticker & \parbox[c]{7cm}{\centering Company} & Adjusted Weight \\
\hline
PG & Procter \& Gamble Co. & 13.34\% \\
KO & Coca-Cola Co. & 10.35\% \\
PM & Philip Morris International Inc. & 8.02\% \\
PEP & PepsiCo Inc. & 7.91\% \\
F & Ford Motor Co. & 4.09\% \\
MO & Altria Group Inc. & 3.85\% \\
KFT & Kraft Foods Inc. Cl A & 3.73\% \\
CL & Colgate-Palmolive Co. & 2.90\% \\
MON & Monsanto Co. & 2.50\% \\
NKE & Nike Inc. Cl B & 1.97\% \\
\hline
\end{tabular}
\end{center}

\subsection{Technology}

\begin{center}\footnotesize
\begin{tabular}{clc}
\hline
ISIN/Ticker & \parbox[c]{7cm}{\centering Company} & Adjusted Weight \\
\hline
AAPL & Apple Inc. & 13.57\% \\
MSFT & Microsoft Corp. & 9.33\% \\
IBM & International Business Machines Corp. & 8.58\% \\
GOOG & Google Inc. Cl A & 6.52\% \\
INTC & Intel Corp. & 5.65\% \\
CSCO & Cisco Systems Inc. & 5.31\% \\
ORCL & Oracle Corp. & 4.98\% \\
HPQ & Hewlett-Packard Co. & 4.73\% \\
QCOM & Qualcomm Inc. & 3.61\% \\
EMC & EMC Corp. & 2.11\% \\
\hline
\end{tabular}
\end{center}

\subsection{Telecommunications}

\begin{center}\footnotesize
\begin{tabular}{clc}
\hline
ISIN/Ticker & \parbox[c]{7cm}{\centering Company} & Adjusted Weight \\
\hline
T & AT\&T Inc. & 44.32\% \\
VZ & Verizon Communications Inc. & 24.29\% \\
AMT & American Tower Corp. Cl A & 5.45\% \\
CTL & CenturyLink Inc. & 3.47\% \\
S & Sprint Nextel Corp. & 2.99\% \\
CCI & Crown Castle International Corp. & 2.75\% \\
Q & \parbox[c]{7cm}{Qwest Communications International Inc.} & 2.67\% \\
FTR & Frontier Communications Corp. & 2.44\% \\
VMED & Virgin Media Inc. & 2.03\% \\
NIHD & NII Holdings Inc. & 1.72\% \\
\hline
\end{tabular}
\end{center}

\subsection{Utilities}

\begin{center}\footnotesize
\begin{tabular}{clc}
\hline
ISIN/Ticker & \parbox[c]{7cm}{\centering Company} & Adjusted Weight \\
\hline
SO & Southern Co. & 6.68\% \\
EXC & Exelon Corp. & 5.61\% \\
D & Dominion Resources Inc. (Virginia) & 5.28\% \\
DUK & Duke Energy Corp. & 4.94\% \\
NEE & NextEra Energy Inc. & 4.52\% \\
PCG & PG\&E Corp. & 3.96\% \\
AEP & American Electric Power Co. Inc. & 3.67\% \\
PEG & Public Service Enterprise Group Inc. & 3.38\% \\
SE & Spectra Energy Corp. & 3.32\% \\
ED & Consolidated Edison Inc. & 2.93\% \\
\hline
\end{tabular}
\end{center}

\section{List of time series segments}
\label{app:listsegments}

In this appendix, we list all time series segments identified by the recursive
segmentation procedure, for all ten DJUS economic sector indices.  In the tables
to follow, the start, end, and duration of each segment is given in terms of the
number of half hours since 14 February 2000.  The actual calendar date for the
start of each segment is also given.

If segment $m$ with $n_m$ half hours is indeed generated by a Gaussian process with
mean $\mu_m$ and standard deviation $\sigma_m$, the standard errors in
estimating $\mu_m$ and $\sigma_m$ are given by the finite-sample formulas
\begin{equation}
\delta\mu_m = \frac{\sigma_m}{\sqrt{n_m}}
\end{equation}
and
\begin{equation}
\delta\sigma_m = \frac{\sigma_m}{\sqrt{2(n_m - 1)}}.
\end{equation}
respectively.  Even if segment $m$ is generated by a different stochastic
process, these formulas are still useful for gauging the magnitudes of the
standard errors in $\mu_m$ and $\sigma_m$.

The Jensen-Shannon divergence $\Delta(m-1, m)$ between successive segments $m-1$
and $m$ are also given.  Given that the Jensen-Shannon divergence between
successive Gaussian segments is a simple function (Eq.~\eqref{eqn:simpleJS}) of
the standard deviations $\sigma_L$ (of segment $m-1$ with length $n_L$),
$\sigma_R$ (of segment $m$ with length $n_R$), and $\sigma$ (of the combined
supersegment with length $n = n_L + n_R$), we estimate the error in $\Delta(m-1,
m)$ as
\begin{equation}
\delta\Delta = n_L\, \frac{\delta\sigma_L}{\sigma_L} + n_R\,
\frac{\delta\sigma_R}{\sigma_R} - n\, \frac{\delta\sigma}{\sigma}.
\end{equation}
Here, we make use of the fact that $\delta\sigma$ is positively correlated to
$\delta\sigma_L$ and $\delta\sigma_R$.

Using the fact that the fractional errors are
\begin{equation}
\frac{\delta\sigma}{\sigma} = \frac{1}{\sqrt{2(n - 1)}}, \quad
\frac{\delta\sigma_L}{\sigma_L} = \frac{1}{\sqrt{2(n_L - 1)}}, \quad
\frac{\delta\sigma_R}{\sigma_R} = \frac{1}{\sqrt{2(n_R - 1)}},
\end{equation}
for Gaussian segments, we arrive at the simplified expression
\begin{equation}\label{eqn:dD}
\delta\Delta = \frac{n_L}{\sqrt{2(n_L - 1)}} + \frac{n_R}{\sqrt{2(n_R - 1)}} -
\frac{n}{\sqrt{2(n-1)}}
\end{equation}
for the error in the Jensen-Shannon divergence $\Delta(m-1, m)$.  

\begin{figure}[htb]
\centering
\includegraphics[scale=0.5,clip=true]{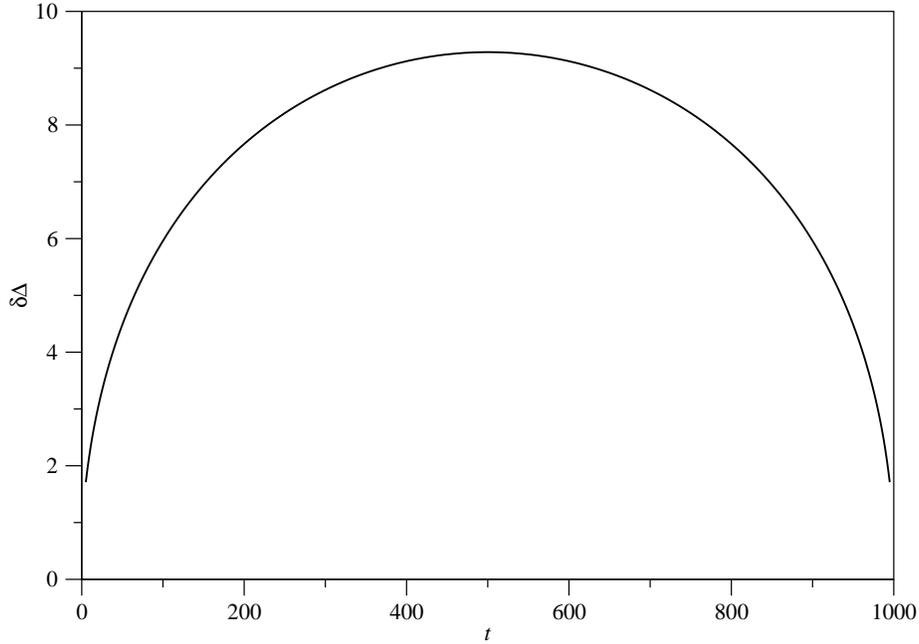}
\caption{Graph of the error $\delta\Delta$ for the Jensen-Shannon divergence of
a Gaussian segment of length $n = 1000$.}
\label{fig:dD1000}
\end{figure}

The error $\delta\Delta$ derived in Eq.~\eqref{eqn:dD} is independent of the
data, and depends only on the position $t$ of the segment boundary.
Fig.~\ref{fig:dD1000} shows $\delta\Delta$ for a segment of length $n = 1000$,
which is largest when $n_L = n_R = n/2$.  This maximum error
$\delta\Delta_{\max}$ grows with the length of the segment as
\begin{equation}\label{eqn:dDmax}
\delta\Delta_{\max} = \sqrt{n}\left(1 - \frac{1}{\sqrt{2}}\right).
\end{equation}

The large majority of our time series segments are shorter than $n = 1000$.
Hence their Jensen-Shannon divergences $\Delta(m-1, m)$ ought to be compared
against standard errors $\delta\Delta$ that are generally smaller than
$\delta\Delta_{\max} = 9.26$ for $n = 1000$.  Moreover, most of the time series
segments longer than $n = 1000$ are enclosed by very strong segment boundaries
with very large Jensen-Shannon divergences.  This suggests that most of our
segments, which are selected based on our empirical cutoff $\Delta_0 = 10$,
should be statistically significant.  

In fact, for shorter segments, Eq.~\eqref{eqn:dDmax} tells us that we can adopt
a cutoff lower than 10, and still maintain their statistical significance.  We
do not do this, because it would result in a large number of short but
statistically significant intraday market microstructure segments.  One of our
reasons for choosing the empirical cutoff of $\Delta_0 = 10$ is to limit the
emergence of such segments.

\subsection{Basic Materials}

\begin{center}\tiny
\begin{tabular}{ccccccccc}
\hline
$m$ & start & end & duration & start date & mean & std dev & $\Delta(m-1, m)$ & cluster \\
\hline
1 & 1 & 184 & 184 & 14/02/2000 & $-0.000282\pm 0.000155$ & $0.002098\pm 0.000110$ & - & high \\ 
2 & 185 & 312 & 128 & 02/03/2000 & $-0.000154\pm 0.000216$ & $0.002448\pm 0.000154$ & $11.0\pm 5.1$ & high \\ 
3 & 313 & 336 & 24 & 14/03/2000 & $0.004323\pm 0.000471$ & $0.002306\pm 0.000340$ & $11.3\pm 2.8$ & extremely high \\ 
4 & 337 & 388 & 52 & 16/03/2000 & $-0.000544\pm 0.000255$ & $0.001839\pm 0.000182$ & $10.7\pm 2.5$ & high \\ 
5 & 389 & 612 & 224 & 22/03/2000 & $0.000025\pm 0.000202$ & $0.003024\pm 0.000143$ & $10.0\pm 4.0$ & high \\ 
6 & 615 & 1196 & 582 & 18/04/2000 & $-0.000197\pm 0.000055$ & $0.001325\pm 0.000039$ & $106.0\pm 7.6$ & low \\ 
7 & 1197 & 1380 & 184 & 26/06/2000 & $0.000165\pm 0.000218$ & $0.002955\pm 0.000154$ & $82.0\pm 7.1$ & high \\
\hline
\end{tabular}
\end{center}

\begin{center}\tiny
\begin{tabular}{ccccccccc}
\hline
$m$ & start & end & duration & start date & mean & std dev & $\Delta(m-1, m)$ & cluster \\
\hline
71 & 18882 & 19047 & 166 & 11/04/2005 & $-0.000430\pm 0.000154$ & $0.001990\pm 0.000110$ & $31.0\pm 4.5$ & high \\ 
72 & 19050 & 19065 & 16 & 26/04/2005 & $-0.000604\pm 0.000231$ & $0.000925\pm 0.000169$ & $45.0\pm 2.4$ & low \\ 
73 & 19066 & 19275 & 210 & 28/04/2005 & $0.000052\pm 0.000095$ & $0.001381\pm 0.000068$ & $16.0\pm 2.5$ & low \\ 
74 & 19276 & 20282 & 1007 & 18/05/2005 & $0.000022\pm 0.000036$ & $0.001151\pm 0.000026$ & $15.1\pm 8.0$ & low \\ 
75 & 20283 & 20721 & 439 & 23/08/2005 & $0.000002\pm 0.000041$ & $0.000852\pm 0.000029$ & $32.0\pm 10.4$ & low \\ 
76 & 20722 & 21001 & 280 & 04/10/2005 & $-0.000006\pm 0.000090$ & $0.001512\pm 0.000064$ & $48.0\pm 7.7$ & moderate \\ 
77 & 21002 & 21943 & 942 & 31/10/2005 & $0.000147\pm 0.000030$ & $0.000928\pm 0.000021$ & $51.0\pm 8.8$ & low \\
\hline
\end{tabular}
\end{center}

\subsection{Consumer Services}

\begin{center}\tiny
\begin{tabular}{ccccccccc}
\hline
$m$ & start & end & duration & start date & mean & std dev & $\Delta(m-1, m)$ & cluster \\
\hline
1 & 1 & 481 & 481 & 14/02/2000 & $0.000159\pm 0.000075$ & $0.001655\pm 0.000053$ & - & moderate \\ 
2 & 482 & 627 & 146 & 04/04/2000 & $-0.000384\pm 0.000252$ & $0.003039\pm 0.000178$ & $43.0\pm 6.4$ & extremely high \\ 
3 & 630 & 700 & 71 & 24/04/2000 & $0.000354\pm 0.000151$ & $0.001275\pm 0.000108$ & $24.0\pm 4.1$ & moderate \\ 
4 & 701 & 717 & 17 & 01/05/2000 & $-0.003661\pm 0.001715$ & $0.007072\pm 0.001250$ & $44.0\pm 2.3$ & extremely high \\ 
5 & 720 & 955 & 236 & 04/05/2000 & $-0.000190\pm 0.000093$ & $0.001426\pm 0.000066$ & $80.0\pm 2.6$ & moderate \\ 
6 & 956 & 972 & 17 & 26/05/2000 & $0.001508\pm 0.001271$ & $0.005239\pm 0.000926$ & $53.0\pm 2.6$ & extremely high \\ 
7 & 975 & 1032 & 58 & 30/05/2000 & $0.000398\pm 0.000267$ & $0.002031\pm 0.000190$ & $16.0\pm 2.2$ & high \\
\hline
\end{tabular}
\end{center}

\begin{center}\tiny
\begin{tabular}{ccccccccc}
\hline
$m$ & start & end & duration & start date & mean & std dev & $\Delta(m-1, m)$ & cluster \\
\hline
21 & 2502 & 2549 & 48 & 08/11/2000 & $-0.001592\pm 0.000418$ & $0.002899\pm 0.000299$ & $50.0\pm 3.4$ & extremely high \\ 
22 & 2551 & 3028 & 478 & 14/11/2000 & $-0.000008\pm 0.000065$ & $0.001417\pm 0.000046$ & $31.0\pm 4.2$ & moderate \\ 
23 & 3029 & 3046 & 18 & 02/01/2001 & $0.004830\pm 0.000986$ & $0.004185\pm 0.000718$ & $48.0\pm 2.8$ & extremely high \\ 
24 & 3047 & 3125 & 79 & 04/01/2001 & $-0.000190\pm 0.000184$ & $0.001631\pm 0.000131$ & $21.0\pm 2.4$ & moderate \\ 
25 & 3126 & 3457 & 332 & 11/01/2001 & $0.000102\pm 0.000059$ & $0.001077\pm 0.000042$ & $12.0\pm 4.9$ & low \\ 
26 & 3458 & 3758 & 301 & 16/02/2001 & $-0.000222\pm 0.000106$ & $0.001835\pm 0.000075$ & $40.0\pm 7.4$ & high \\ 
27 & 3759 & 3774 & 16 & 20/03/2001 & $-0.005422\pm 0.000975$ & $0.003901\pm 0.000712$ & $24.0\pm 2.6$ & extremely high \\
\hline
\end{tabular}
\end{center}

\begin{center}\tiny
\begin{tabular}{ccccccccc}
\hline
$m$ & start & end & duration & start date & mean & std dev & $\Delta(m-1, m)$ & cluster \\
\hline
91 & 25112 & 25274 & 163 & 05/12/2006 & $0.000076\pm 0.000049$ & $0.000624\pm 0.000035$ & $40.0\pm 7.5$ & extremely low \\ 
92 & 25276 & 25375 & 100 & 20/12/2006 & $-0.000021\pm 0.000041$ & $0.000405\pm 0.000029$ & $11.1\pm 4.7$ & extremely low \\ 
93 & 25376 & 25933 & 558 & 29/12/2006 & $0.000069\pm 0.000026$ & $0.000618\pm 0.000019$ & $32.9\pm 5.7$ & extremely low \\ 
94 & 25934 & 25966 & 33 & 26/02/2007 & $-0.001285\pm 0.000424$ & $0.002437\pm 0.000305$ & $138.0\pm 3.6$ & high \\ 
95 & 25967 & 26310 & 344 & 01/03/2007 & $0.000096\pm 0.000052$ & $0.000968\pm 0.000037$ & $37.0\pm 3.5$ & low \\ 
96 & 26311 & 26381 & 71 & 03/04/2007 & $0.000071\pm 0.000035$ & $0.000297\pm 0.000025$ & $44.0\pm 4.7$ & extremely low \\ 
97 & 26385 & 26644 & 260 & 10/04/2007 & $0.000022\pm 0.000048$ & $0.000769\pm 0.000034$ & $27.0\pm 4.5$ & low \\
\hline
\end{tabular}
\end{center}

\subsection{Oil \& Gas}

\begin{center}\tiny
\begin{tabular}{ccccccccc}
\hline
$m$ & start & end & duration & start date & mean & std dev & $\Delta(m-1, m)$ & cluster \\
\hline
1 & 1 & 223 & 223 & 14/02/2000 & $0.000302\pm 0.000138$ & $0.002055\pm 0.000098$ & - & moderate \\ 
2 & 224 & 251 & 28 & 06/03/2000 & $0.002718\pm 0.000959$ & $0.005074\pm 0.000690$ & $28.0\pm 3.2$ & extremely high \\ 
3 & 256 & 616 & 361 & 09/03/2000 & $0.000022\pm 0.000136$ & $0.002587\pm 0.000096$ & $15.0\pm 3.2$ & high \\ 
4 & 617 & 1001 & 385 & 24/04/2000 & $0.000082\pm 0.000067$ & $0.001307\pm 0.000047$ & $83.0\pm 8.0$ & moderate \\ 
5 & 1002 & 1167 & 166 & 05/06/2000 & $0.000254\pm 0.000177$ & $0.002283\pm 0.000126$ & $37.0\pm 6.4$ & high \\ 
6 & 1170 & 1251 & 82 & 22/06/2000 & $-0.000217\pm 0.000144$ & $0.001301\pm 0.000102$ & $13.6\pm 4.4$ & moderate \\ 
7 & 1252 & 1335 & 84 & 30/06/2000 & $-0.000225\pm 0.000330$ & $0.003022\pm 0.000235$ & $18.2\pm 3.8$ & high \\
\hline
\end{tabular}
\end{center}

\begin{center}\tiny
\begin{tabular}{ccccccccc}
\hline
$m$ & start & end & duration & start date & mean & std dev & $\Delta(m-1, m)$ & cluster \\
\hline
41 & 7797 & 8531 & 735 & 03/05/2002 & $-0.000236\pm 0.000060$ & $0.001622\pm 0.000042$ & $12.0\pm 7.4$ & moderate \\ 
42 & 8532 & 8772 & 241 & 12/07/2002 & $-0.000587\pm 0.000235$ & $0.003656\pm 0.000167$ & $122.0\pm 8.1$ & extremely high \\ 
43 & 8775 & 9676 & 902 & 05/08/2002 & $0.000068\pm 0.000084$ & $0.002519\pm 0.000059$ & $26.0\pm 8.3$ & high \\ 
44 & 9677 & 10343 & 667 & 30/10/2002 & $0.000100\pm 0.000053$ & $0.001379\pm 0.000038$ & $112.0\pm 11.5$ & moderate \\ 
45 & 10344 & 10381 & 38 & 06/01/2003 & $-0.000827\pm 0.000329$ & $0.002025\pm 0.000235$ & $10.6\pm 3.9$ & moderate \\ 
46 & 10382 & 10452 & 71 & 09/01/2003 & $-0.000043\pm 0.000105$ & $0.000887\pm 0.000075$ & $11.0\pm 3.0$ & low \\ 
47 & 10453 & 10469 & 17 & 15/01/2003 & $0.000529\pm 0.001404$ & $0.005787\pm 0.001023$ & $57.0\pm 2.3$ & extremely high \\
\hline
\end{tabular}
\end{center}

\subsection{Financials}

\begin{center}\tiny
\begin{tabular}{ccccccccc}
\hline
$m$ & start & end & duration & start date & mean & std dev & $\Delta(m-1, m)$ & cluster \\
\hline
1 & 1 & 316 & 316 & 14/02/2000 & $-0.000181\pm 0.000088$ & $0.001564\pm 0.000062$ & - & moderate \\ 
2 & 317 & 702 & 386 & 15/03/2000 & $0.000407\pm 0.000133$ & $0.002606\pm 0.000094$ & $39.0\pm 7.8$ & high \\ 
3 & 705 & 731 & 27 & 01/05/2000 & $-0.001205\pm 0.000120$ & $0.000623\pm 0.000086$ & $20.0\pm 3.2$ & low \\ 
4 & 732 & 970 & 239 & 05/05/2000 & $0.000261\pm 0.000119$ & $0.001840\pm 0.000084$ & $15.0\pm 3.1$ & high \\ 
5 & 971 & 987 & 17 & 30/05/2000 & $0.004637\pm 0.001990$ & $0.008205\pm 0.001450$ & $63.0\pm 2.6$ & extremely high \\ 
6 & 990 & 1076 & 87 & 02/06/2000 & $-0.000365\pm 0.000202$ & $0.001887\pm 0.000144$ & $25.0\pm 2.3$ & high \\ 
7 & 1077 & 1092 & 16 & 14/06/2000 & $-0.001703\pm 0.001245$ & $0.004979\pm 0.000909$ & $11.0\pm 2.3$ & extremely high \\
\hline
\end{tabular}
\end{center}

\begin{center}\tiny
\begin{tabular}{ccccccccc}
\hline
$m$ & start & end & duration & start date & mean & std dev & $\Delta(m-1, m)$ & cluster \\
\hline
71 & 16920 & 17021 & 102 & 01/10/2004 & $-0.000073\pm 0.000048$ & $0.000484\pm 0.000034$ & $15.0\pm 4.2$ & low \\ 
72 & 17022 & 17142 & 121 & 13/10/2004 & $-0.000324\pm 0.000107$ & $0.001176\pm 0.000076$ & $28.0\pm 4.4$ & moderate \\ 
73 & 17145 & 17270 & 126 & 25/10/2004 & $0.000485\pm 0.000070$ & $0.000784\pm 0.000050$ & $43.0\pm 4.6$ & low \\ 
74 & 17271 & 17845 & 575 & 05/11/2004 & $0.000092\pm 0.000025$ & $0.000606\pm 0.000018$ & $19.0\pm 6.2$ & low \\ 
75 & 17846 & 18147 & 302 & 31/12/2004 & $-0.000119\pm 0.000044$ & $0.000760\pm 0.000031$ & $20.0\pm 8.3$ & low \\ 
76 & 18148 & 18313 & 166 & 31/01/2005 & $0.000150\pm 0.000039$ & $0.000509\pm 0.000028$ & $17.0\pm 6.1$ & low \\ 
77 & 18314 & 18433 & 120 & 15/02/2005 & $-0.000232\pm 0.000064$ & $0.000705\pm 0.000046$ & $10.0\pm 4.9$ & low \\
\hline
\end{tabular}
\end{center}

\subsection{Healthcare}

\begin{center}\tiny
\begin{tabular}{ccccccccc}
\hline
$m$ & start & end & duration & start date & mean & std dev & $\Delta(m-1, m)$ & cluster \\
\hline
1 & 1 & 73 & 73 & 14/02/2000 & $0.000123\pm 0.000147$ & $0.001258\pm 0.000105$ & - & high \\ 
2 & 74 & 133 & 60 & 18/02/2000 & $-0.003154\pm 0.002585$ & $0.020025\pm 0.001843$ & $143.0\pm 3.4$ & extremely high \\ 
3 & 136 & 149 & 14 & 28/02/2000 & $0.012404\pm 0.000489$ & $0.001829\pm 0.000359$ & $2165.0\pm 2.1$ & high \\ 
4 & 151 & 223 & 73 & 29/02/2000 & $0.000016\pm 0.000141$ & $0.001203\pm 0.000100$ & $118.0\pm 2.2$ & moderate \\ 
5 & 224 & 495 & 272 & 06/03/2000 & $0.000225\pm 0.000177$ & $0.002916\pm 0.000125$ & $27.0\pm 4.6$ & very high \\ 
6 & 496 & 544 & 49 & 05/04/2000 & $0.000443\pm 0.000153$ & $0.001074\pm 0.000110$ & $106.0\pm 4.0$ & moderate \\ 
7 & 545 & 601 & 57 & 12/04/2000 & $-0.000760\pm 0.000350$ & $0.002645\pm 0.000250$ & $11.2\pm 3.1$ & very high \\ 
8 & 602 & 1525 & 924 & 18/04/2000 & $0.000126\pm 0.000049$ & $0.001504\pm 0.000035$ & $23.1\pm 4.7$ & high \\ 
9 & 1526 & 1542 & 17 & 31/07/2000 & $0.002920\pm 0.001327$ & $0.005472\pm 0.000967$ & $30.3\pm 2.8$ & extremely high \\ 
10 & 1545 & 1710 & 166 & 03/08/2000 & $-0.000292\pm 0.000111$ & $0.001428\pm 0.000079$ & $45.4\pm 2.5$ & high \\ 
11 & 1711 & 1825 & 115 & 21/08/2000 & $0.000102\pm 0.000069$ & $0.000737\pm 0.000049$ & $25.0\pm 4.9$ & low \\ 
12 & 1826 & 1842 & 17 & 30/08/2000 & $-0.002606\pm 0.000849$ & $0.003502\pm 0.000619$ & $52.2\pm 2.5$ & very high \\ 
13 & 1845 & 1986 & 142 & 06/09/2000 & $0.000129\pm 0.000073$ & $0.000873\pm 0.000052$ & $46.1\pm 2.5$ & moderate \\ 
14 & 1987 & 3031 & 1045 & 20/09/2000 & $0.000119\pm 0.000050$ & $0.001602\pm 0.000035$ & $21.1\pm 7.0$ & high \\ 
15 & 3032 & 3091 & 60 & 03/01/2001 & $-0.001429\pm 0.000442$ & $0.003427\pm 0.000315$ & $49.0\pm 4.9$ & very high \\ 
\hline
\end{tabular}
\end{center}

\begin{center}\tiny
\begin{tabular}{ccccccccc}
\hline
$m$ & start & end & duration & start date & mean & std dev & $\Delta(m-1, m)$ & cluster \\
\hline
16 & 3092 & 3208 & 117 & 09/01/2001 & $-0.000100\pm 0.000115$ & $0.001242\pm 0.000082$ & $39.0\pm 3.8$ & high \\ 
17 & 3209 & 3239 & 31 & 19/01/2001 & $0.000896\pm 0.000444$ & $0.002471\pm 0.000319$ & $15.0\pm 3.1$ & very high \\ 
18 & 3241 & 3256 & 16 & 26/01/2001 & $0.001508\pm 0.000314$ & $0.001257\pm 0.000230$ & $27.0\pm 2.0$ & high \\ 
19 & 3257 & 3628 & 372 & 29/01/2001 & $-0.000134\pm 0.000060$ & $0.001158\pm 0.000043$ & $19.0\pm 2.6$ & moderate \\ 
20 & 3629 & 3761 & 133 & 06/03/2001 & $-0.000705\pm 0.000171$ & $0.001968\pm 0.000121$ & $21.0\pm 5.9$ & high \\ 
21 & 3762 & 3777 & 16 & 20/03/2001 & $-0.004733\pm 0.002039$ & $0.008155\pm 0.001489$ & $48.0\pm 2.4$ & extremely high \\ 
22 & 3778 & 4171 & 394 & 22/03/2001 & $0.000326\pm 0.000093$ & $0.001847\pm 0.000066$ & $77.0\pm 2.6$ & high \\
\hline
\end{tabular}
\end{center}

\begin{center}\tiny
\begin{tabular}{ccccccccc}
\hline
$m$ & start & end & duration & start date & mean & std dev & $\Delta(m-1, m)$ & cluster \\
\hline
91 & 22813 & 22974 & 162 & 27/04/2006 & $-0.000069\pm 0.000039$ & $0.000491\pm 0.000027$ & $17.0\pm 7.7$ & low \\ 
92 & 22975 & 23632 & 658 & 12/05/2006 & $-0.000015\pm 0.000033$ & $0.000856\pm 0.000024$ & $20.0\pm 6.9$ & moderate \\ 
93 & 23633 & 23650 & 18 & 17/07/2006 & $-0.000424\pm 0.000517$ & $0.002193\pm 0.000376$ & $25.0\pm 2.8$ & very high \\ 
94 & 23653 & 23815 & 163 & 18/07/2006 & $0.000350\pm 0.000073$ & $0.000930\pm 0.000052$ & $14.0\pm 2.6$ & moderate \\ 
95 & 23818 & 25471 & 1654 & 02/08/2006 & $0.000036\pm 0.000016$ & $0.000651\pm 0.000011$ & $49.0\pm 7.7$ & low \\ 
96 & 25472 & 25931 & 460 & 11/01/2007 & $0.000036\pm 0.000023$ & $0.000486\pm 0.000016$ & $27.0\pm 11.4$ & low \\ 
97 & 25932 & 25965 & 34 & 26/02/2007 & $-0.000710\pm 0.000342$ & $0.001994\pm 0.000245$ & $121.0\pm 3.6$ & high \\
\hline
\end{tabular}
\end{center}

\subsection{Industrials}

\begin{center}\tiny
\begin{tabular}{ccccccccc}
\hline
$m$ & start & end & duration & start date & mean & std dev & $\Delta(m-1, m)$ & cluster \\
\hline
1 & 1 & 67 & 67 & 14/02/2000 & $0.000031\pm 0.000144$ & $0.001179\pm 0.000103$ & - & moderate \\ 
2 & 68 & 284 & 217 & 18/02/2000 & $0.000138\pm 0.000139$ & $0.002051\pm 0.000099$ & $17.0\pm 4.3$ & high \\ 
3 & 285 & 482 & 198 & 10/03/2000 & $0.000027\pm 0.000178$ & $0.002509\pm 0.000126$ & $10.7\pm 6.0$ & high \\ 
4 & 483 & 602 & 120 & 04/04/2000 & $-0.000240\pm 0.000366$ & $0.004006\pm 0.000260$ & $31.0\pm 5.1$ & extremely high \\ 
5 & 603 & 701 & 99 & 18/04/2000 & $0.000290\pm 0.000193$ & $0.001921\pm 0.000137$ & $30.0\pm 4.4$ & high \\ 
6 & 702 & 720 & 19 & 01/05/2000 & $-0.000589\pm 0.001123$ & $0.004894\pm 0.000816$ & $21.9\pm 2.5$ & extremely high \\ 
7 & 721 & 972 & 252 & 05/05/2000 & $-0.000053\pm 0.000131$ & $0.002077\pm 0.000093$ & $15.4\pm 2.8$ & high \\
\hline
\end{tabular}
\end{center}

\begin{center}\tiny
\begin{tabular}{ccccccccc}
\hline
$m$ & start & end & duration & start date & mean & std dev & $\Delta(m-1, m)$ & cluster \\
\hline
41 & 6428 & 6537 & 110 & 19/12/2001 & $0.000076\pm 0.000061$ & $0.000644\pm 0.000044$ & $26.7\pm 5.7$ & low \\ 
42 & 6538 & 6702 & 165 & 31/12/2001 & $-0.000180\pm 0.000086$ & $0.001107\pm 0.000061$ & $15.9\pm 4.8$ & moderate \\ 
43 & 6703 & 6943 & 241 & 16/01/2002 & $-0.000060\pm 0.000114$ & $0.001764\pm 0.000081$ & $10.3\pm 5.8$ & high \\ 
44 & 6944 & 7823 & 880 & 08/02/2002 & $-0.000017\pm 0.000046$ & $0.001355\pm 0.000032$ & $9.9\pm 8.3$ & moderate \\ 
45 & 7824 & 7845 & 22 & 07/05/2002 & $0.000660\pm 0.000736$ & $0.003453\pm 0.000533$ & $24.3\pm 3.1$ & extremely high \\ 
46 & 7846 & 8082 & 237 & 09/05/2002 & $-0.000026\pm 0.000083$ & $0.001279\pm 0.000059$ & $23.1\pm 2.9$ & moderate \\ 
47 & 8083 & 8395 & 313 & 31/05/2002 & $-0.000133\pm 0.000102$ & $0.001811\pm 0.000072$ & $11.4\pm 6.8$ & high \\
\hline
\end{tabular}
\end{center}

\begin{center}\tiny
\begin{tabular}{ccccccccc}
\hline
$m$ & start & end & duration & start date & mean & std dev & $\Delta(m-1, m)$ & cluster \\
\hline
111 & 29425 & 29832 & 408 & 30/01/2008 & $-0.000061\pm 0.000075$ & $0.001518\pm 0.000053$ & $11.3\pm 4.1$ & moderate \\ 
112 & 29833 & 29955 & 123 & 10/03/2008 & $0.000290\pm 0.000213$ & $0.002368\pm 0.000152$ & $23.0\pm 5.9$ & high \\ 
113 & 29956 & 30152 & 197 & 24/03/2008 & $0.000049\pm 0.000083$ & $0.001168\pm 0.000059$ & $34.0\pm 5.2$ & moderate \\ 
114 & 30153 & 30169 & 17 & 10/04/2008 & $-0.000946\pm 0.000867$ & $0.003577\pm 0.000632$ & $27.0\pm 2.6$ & extremely high \\ 
115 & 30170 & 30207 & 38 & 11/04/2008 & $-0.000042\pm 0.000127$ & $0.000785\pm 0.000091$ & $24.0\pm 2.1$ & low \\ 
116 & 30208 & 30255 & 48 & 15/04/2008 & $0.000408\pm 0.000292$ & $0.002023\pm 0.000209$ & $13.0\pm 2.8$ & high \\ 
117 & 30256 & 30576 & 321 & 21/04/2008 & $0.000089\pm 0.000056$ & $0.000997\pm 0.000039$ & $45.0\pm 4.0$ & moderate \\ 
118 & 30577 & 30632 & 56 & 19/05/2008 & $-0.000492\pm 0.000186$ & $0.001394\pm 0.000133$ & $11.0\pm 4.3$ & moderate \\ 
119 & 30633 & 30747 & 115 & 23/05/2008 & $0.000137\pm 0.000096$ & $0.001028\pm 0.000068$ & $12.0\pm 3.7$ & moderate \\ 
120 & 30748 & 30825 & 78 & 05/06/2008 & $-0.000272\pm 0.000240$ & $0.002119\pm 0.000171$ & $30.0\pm 4.1$ & high \\ 
121 & 30826 & 31026 & 201 & 13/06/2008 & $-0.000249\pm 0.000100$ & $0.001424\pm 0.000071$ & $7.9\pm 4.5$ & moderate \\ 
122 & 31027 & 31097 & 71 & 03/07/2008 & $-0.000142\pm 0.000377$ & $0.003177\pm 0.000268$ & $24.0\pm 4.4$ & extremely high \\ 
123 & 31098 & 31557 & 460 & 15/07/2008 & $0.000079\pm 0.000072$ & $0.001537\pm 0.000051$ & $35.0\pm 4.9$ & moderate \\ 
\hline
\end{tabular}
\end{center}

\subsection{Consumer Goods}

\begin{center}\tiny
\begin{tabular}{ccccccccc}
\hline
$m$ & start & end & duration & start date & mean & std dev & $\Delta(m-1, m)$ & cluster \\
\hline
1 & 1 & 211 & 211 & 14/02/2000 & $-0.000082\pm 0.000104$ & $0.001517\pm 0.000074$ & - & high \\ 
2 & 212 & 226 & 15 & 06/03/2000 & $-0.000548\pm 0.000285$ & $0.001103\pm 0.000209$ & $42.0\pm 2.5$ & high \\ 
3 & 227 & 244 & 18 & 07/03/2000 & $-0.001438\pm 0.001211$ & $0.005139\pm 0.000881$ & $14.0\pm 1.8$ & extremely high \\ 
4 & 245 & 473 & 229 & 08/03/2000 & $0.000190\pm 0.000135$ & $0.002041\pm 0.000096$ & $15.0\pm 2.7$ & extremely high \\ 
5 & 474 & 490 & 17 & 03/04/2000 & $0.000447\pm 0.001368$ & $0.005639\pm 0.000997$ & $23.0\pm 2.6$ & extremely high \\ 
6 & 491 & 602 & 112 & 04/04/2000 & $-0.000361\pm 0.000253$ & $0.002675\pm 0.000180$ & $10.7\pm 2.5$ & extremely high \\ 
7 & 603 & 676 & 74 & 18/04/2000 & $0.000043\pm 0.000204$ & $0.001759\pm 0.000146$ & $13.0\pm 4.0$ & extremely high \\
\hline
\end{tabular}
\end{center}

\begin{center}\tiny
\begin{tabular}{ccccccccc}
\hline
$m$ & start & end & duration & start date & mean & std dev & $\Delta(m-1, m)$ & cluster \\
\hline
61 & 12550 & 12702 & 153 & 06/08/2003 & $0.000034\pm 0.000044$ & $0.000544\pm 0.000031$ & $37.0\pm 6.2$ & low \\ 
62 & 12703 & 13727 & 1025 & 20/08/2003 & $0.000042\pm 0.000023$ & $0.000740\pm 0.000016$ & $8.0\pm 7.1$ & moderate \\ 
63 & 13728 & 13877 & 150 & 24/11/2003 & $0.000020\pm 0.000048$ & $0.000587\pm 0.000034$ & $21.0\pm 7.1$ & low \\ 
64 & 13878 & 14774 & 897 & 09/12/2003 & $0.000047\pm 0.000019$ & $0.000581\pm 0.000014$ & $11.0\pm 7.0$ & low \\ 
65 & 14775 & 14896 & 122 & 09/03/2004 & $-0.000130\pm 0.000093$ & $0.001024\pm 0.000066$ & $45.0\pm 6.4$ & high \\ 
66 & 14897 & 14911 & 15 & 22/03/2004 & $-0.000065\pm 0.000155$ & $0.000601\pm 0.000114$ & $13.0\pm 2.4$ & low \\ 
67 & 14912 & 15203 & 292 & 23/03/2004 & $0.000093\pm 0.000035$ & $0.000599\pm 0.000025$ & $14.0\pm 2.5$ & low \\
\hline
\end{tabular}
\end{center}

\subsection{Technology}

\begin{center}\tiny
\begin{tabular}{ccccccccc}
\hline
$m$ & start & end & duration & start date & mean & std dev & $\Delta(m-1, m)$ & cluster \\
\hline
1 & 1 & 282 & 282 & 14/02/2000 & $0.000290\pm 0.000147$ & $0.002461\pm 0.000147$ & - & high \\ 
2 & 283 & 464 & 182 & 10/03/2000 & $-0.000166\pm 0.000290$ & $0.003916\pm 0.000291$ & $23.0\pm 6.2$ & very high \\ 
3 & 465 & 630 & 166 & 31/03/2000 & $-0.000530\pm 0.000608$ & $0.007838\pm 0.000610$ & $34.0\pm 5.5$ & extremely high \\ 
4 & 631 & 971 & 341 & 25/04/2000 & $-0.000057\pm 0.000228$ & $0.004211\pm 0.000228$ & $41.0\pm 6.3$ & very high \\ 
5 & 972 & 990 & 19 & 30/05/2000 & $0.002926\pm 0.002237$ & $0.009752\pm 0.002299$ & $17.0\pm 2.8$ & extremely high \\ 
6 & 991 & 1482 & 492 & 05/06/2000 & $0.000013\pm 0.000101$ & $0.002232\pm 0.000101$ & $87.0\pm 2.9$ & high \\ 
7 & 1483 & 1546 & 64 & 25/07/2000 & $-0.000514\pm 0.000602$ & $0.004820\pm 0.000607$ & $42.0\pm 4.7$ & very high \\
\hline
\end{tabular}
\end{center}

\begin{center}\tiny
\begin{tabular}{ccccccccc}
\hline
$m$ & start & end & duration & start date & mean & std dev & $\Delta(m-1, m)$ & cluster \\
\hline
71 & 16347 & 16361 & 15 & 09/08/2004 & $0.000547\pm 0.000326$ & $0.001261\pm 0.000337$ & $34.9\pm 2.6$ & moderate \\ 
72 & 16362 & 16380 & 19 & 10/08/2004 & $-0.001493\pm 0.001085$ & $0.004730\pm 0.001115$ & $53.0\pm 1.8$ & extremely high \\ 
73 & 16381 & 16396 & 16 & 12/08/2004 & $-0.000009\pm 0.000308$ & $0.001232\pm 0.000318$ & $65.0\pm 1.8$ & low \\ 
74 & 16397 & 16572 & 176 & 13/08/2004 & $0.000118\pm 0.000081$ & $0.001078\pm 0.000082$ & $14.0\pm 2.5$ & low \\ 
75 & 16573 & 16695 & 123 & 30/08/2004 & $0.000184\pm 0.000153$ & $0.001699\pm 0.000154$ & $17.3\pm 5.0$ & moderate \\ 
76 & 16696 & 16901 & 206 & 13/09/2004 & $-0.000038\pm 0.000077$ & $0.001100\pm 0.000077$ & $13.7\pm 5.2$ & low \\ 
77 & 16902 & 16920 & 19 & 30/09/2004 & $0.001335\pm 0.000611$ & $0.002663\pm 0.000628$ & $16.0\pm 2.7$ & high \\
\hline
\end{tabular}
\end{center}

\subsection{Telecommunications}

\begin{center}\tiny
\begin{tabular}{ccccccccc}
\hline
$m$ & start & end & duration & start date & mean & std dev & $\Delta(m-1, m)$ & cluster \\
\hline
1 & 1 & 162 & 162 & 14/02/2000 & $-0.000146\pm 0.000127$ & $0.001617\pm 0.000090$ & - & moderate \\ 
2 & 163 & 183 & 21 & 29/02/2000 & $0.001959\pm 0.000755$ & $0.003462\pm 0.000547$ & $23.1\pm 2.8$ & extremely high \\ 
3 & 184 & 477 & 294 & 02/03/2000 & $0.000036\pm 0.000099$ & $0.001705\pm 0.000070$ & $20.9\pm 2.9$ & moderate \\ 
4 & 478 & 630 & 153 & 03/04/2000 & $-0.000366\pm 0.000262$ & $0.003245\pm 0.000186$ & $39.0\pm 6.0$ & extremely high \\ 
5 & 631 & 701 & 71 & 25/04/2000 & $0.000240\pm 0.000189$ & $0.001591\pm 0.000134$ & $18.0\pm 4.2$ & moderate \\ 
6 & 702 & 720 & 19 & 01/05/2000 & $-0.001939\pm 0.002107$ & $0.009186\pm 0.001531$ & $46.0\pm 2.4$ & extremely high \\ 
7 & 721 & 867 & 147 & 05/05/2000 & $-0.000099\pm 0.000135$ & $0.001639\pm 0.000096$ & $75.0\pm 2.6$ & moderate \\ 
8 & 868 & 956 & 89 & 18/05/2000 & $-0.000363\pm 0.000283$ & $0.002666\pm 0.000201$ & $24.0\pm 4.4$ & high \\ 
9 & 957 & 975 & 19 & 26/05/2000 & $0.002296\pm 0.001156$ & $0.005039\pm 0.000840$ & $10.0\pm 2.5$ & extremely high \\ 
10 & 976 & 991 & 16 & 02/06/2000 & $0.000458\pm 0.000244$ & $0.000976\pm 0.000178$ & $80.0\pm 1.8$ & low \\ 
11 & 992 & 1482 & 491 & 05/06/2000 & $-0.000033\pm 0.000068$ & $0.001502\pm 0.000048$ & $41.0\pm 2.7$ & moderate \\ 
12 & 1483 & 1515 & 33 & 25/07/2000 & $-0.001291\pm 0.000542$ & $0.003116\pm 0.000389$ & $27.2\pm 3.6$ & extremely high \\ 
13 & 1516 & 2067 & 552 & 31/07/2000 & $-0.000116\pm 0.000055$ & $0.001298\pm 0.000039$ & $39.4\pm 3.6$ & moderate \\ 
14 & 2068 & 2433 & 366 & 27/09/2000 & $0.000049\pm 0.000138$ & $0.002638\pm 0.000098$ & $95.0\pm 8.7$ & high \\ 
15 & 2434 & 2582 & 149 & 02/11/2000 & $-0.000048\pm 0.000109$ & $0.001332\pm 0.000077$ & $44.9\pm 6.1$ & moderate \\ 
\hline
\end{tabular}
\end{center}

\begin{center}\tiny
\begin{tabular}{ccccccccc}
\hline
$m$ & start & end & duration & start date & mean & std dev & $\Delta(m-1, m)$ & cluster \\
\hline
16 & 2583 & 2598 & 16 & 16/11/2000 & $-0.001410\pm 0.001237$ & $0.004950\pm 0.000904$ & $41.7\pm 2.5$ & extremely high \\ 
17 & 2599 & 2879 & 281 & 17/11/2000 & $0.000044\pm 0.000102$ & $0.001709\pm 0.000072$ & $30.9\pm 2.6$ & moderate \\ 
18 & 2880 & 2934 & 55 & 14/12/2000 & $-0.001567\pm 0.000516$ & $0.003825\pm 0.000368$ & $46.9\pm 4.2$ & extremely high \\ 
19 & 2935 & 3029 & 95 & 21/12/2000 & $0.000322\pm 0.000153$ & $0.001493\pm 0.000109$ & $37.1\pm 3.5$ & moderate \\ 
20 & 3030 & 3047 & 18 & 02/01/2001 & $0.002771\pm 0.001272$ & $0.005398\pm 0.000926$ & $38.0\pm 2.5$ & extremely high \\ 
21 & 3048 & 3228 & 181 & 04/01/2001 & $0.000047\pm 0.000165$ & $0.002219\pm 0.000117$ & $23.7\pm 2.6$ & high \\ 
22 & 3229 & 3244 & 16 & 23/01/2001 & $-0.001067\pm 0.001410$ & $0.005639\pm 0.001029$ & $18.7\pm 2.5$ & extremely high \\
\hline
\end{tabular}
\end{center}

\begin{center}\tiny
\begin{tabular}{ccccccccc}
\hline
$m$ & start & end & duration & start date & mean & std dev & $\Delta(m-1, m)$ & cluster \\
\hline
91 & 25624 & 25941 & 318 & 26/01/2007 & $0.000049\pm 0.000039$ & $0.000691\pm 0.000027$ & $47.7\pm 3.5$ & low \\ 
92 & 25942 & 25967 & 26 & 27/02/2007 & $-0.000219\pm 0.000788$ & $0.004016\pm 0.000568$ & $151.0\pm 3.2$ & extremely high \\ 
93 & 25968 & 26042 & 75 & 01/03/2007 & $-0.000095\pm 0.000170$ & $0.001473\pm 0.000121$ & $21.0\pm 2.7$ & moderate \\ 
94 & 26043 & 26268 & 226 & 08/03/2007 & $0.000086\pm 0.000070$ & $0.001045\pm 0.000049$ & $35.0\pm 4.5$ & low \\ 
95 & 26269 & 26952 & 684 & 29/03/2007 & $0.000074\pm 0.000033$ & $0.000862\pm 0.000023$ & $14.8\pm 7.8$ & low \\ 
96 & 26953 & 27075 & 123 & 04/06/2007 & $-0.000033\pm 0.000112$ & $0.001242\pm 0.000079$ & $12.0\pm 6.3$ & moderate \\ 
97 & 27076 & 27243 & 168 & 15/06/2007 & $-0.000012\pm 0.000087$ & $0.001132\pm 0.000062$ & $13.5\pm 5.0$ & moderate \\
\hline
\end{tabular}
\end{center}

\subsection{Utilities}

\begin{center}\tiny
\begin{tabular}{ccccccccc}
\hline
$m$ & start & end & duration & start date & mean & std dev & $\Delta(m-1, m)$ & cluster \\
\hline
1 & 1 & 74 & 74 & 14/02/2000 & $-0.000042\pm 0.000083$ & $0.000715\pm 0.000059$ & - & low \\ 
2 & 75 & 313 & 239 & 18/02/2000 & $-0.000132\pm 0.000081$ & $0.001245\pm 0.000057$ & $13.2\pm 4.5$ & high \\ 
3 & 314 & 336 & 23 & 14/03/2000 & $0.001750\pm 0.000314$ & $0.001507\pm 0.000227$ & $36.0\pm 3.0$ & extremely high \\ 
4 & 337 & 482 & 146 & 16/03/2000 & $0.000057\pm 0.000110$ & $0.001329\pm 0.000078$ & $11.8\pm 2.8$ & high \\ 
5 & 483 & 615 & 133 & 04/04/2000 & $0.000210\pm 0.000239$ & $0.002752\pm 0.000169$ & $28.4\pm 4.9$ & extremely high \\ 
6 & 616 & 631 & 16 & 24/04/2000 & $0.000176\pm 0.000232$ & $0.000929\pm 0.000170$ & $30.4\pm 2.4$ & low \\ 
7 & 632 & 1080 & 449 & 25/04/2000 & $0.000041\pm 0.000081$ & $0.001716\pm 0.000057$ & $20.8\pm 2.7$ & high \\
\hline
\end{tabular}
\end{center}

\begin{center}\tiny
\begin{tabular}{ccccccccc}
\hline
$m$ & start & end & duration & start date & mean & std dev & $\Delta(m-1, m)$ & cluster \\
\hline
41 & 5593 & 5625 & 33 & 25/09/2001 & $-0.000428\pm 0.000709$ & $0.004073\pm 0.000509$ & $27.9\pm 2.5$ & extremely high \\ 
42 & 5626 & 5687 & 62 & 28/09/2001 & $0.000584\pm 0.000208$ & $0.001639\pm 0.000148$ & $43.4\pm 2.8$ & high \\ 
43 & 5688 & 6242 & 555 & 04/10/2001 & $-0.000117\pm 0.000066$ & $0.001563\pm 0.000047$ & $10.7\pm 4.7$ & high \\ 
44 & 6243 & 6344 & 102 & 29/11/2001 & $0.000072\pm 0.000111$ & $0.001124\pm 0.000079$ & $10.1\pm 5.7$ & moderate \\ 
45 & 6345 & 6362 & 18 & 07/12/2001 & $-0.000783\pm 0.001526$ & $0.006475\pm 0.001111$ & $82.0\pm 2.5$ & extremely high \\ 
46 & 6363 & 6558 & 196 & 13/12/2001 & $0.000126\pm 0.000096$ & $0.001344\pm 0.000068$ & $202.0\pm 2.6$ & high \\ 
47 & 6559 & 6674 & 116 & 03/01/2002 & $-0.000047\pm 0.000076$ & $0.000813\pm 0.000054$ & $23.0\pm 5.1$ & low \\
\hline
\end{tabular}
\end{center}

\end{document}